\documentclass[11pt]{article}
\usepackage{amsmath}
\usepackage{graphicx,psfrag,epsf}
\usepackage{enumerate}
\usepackage{natbib}
\usepackage{amstext}
\usepackage{amssymb}
\usepackage{multirow}
\usepackage{color}
\usepackage{bigdelim}
\usepackage{xr}
\usepackage{soul}

\newcommand{\blind}{0}

\begin{document}

\def\spacingset#1{\renewcommand{\baselinestretch}%
{#1}\small\normalsize} \spacingset{1}
\if0\blind
{
  \title{\bf Estimating trends in the global mean temperature record}
  \author{Andrew Poppick\\
%  \thanks{
%The authors thank Jonah Bloch-Johnson, Malte Jansen, Cristian Proistosescu, and Kate Marvel for helpful conversations and comments related to parts of this work. We additionally thank the reviewers of this paper, whose suggestions led to a number of improvements. This work was supported in part by STATMOS, the Research Network for Statistical Methods for Atmospheric and Oceanic Sciences (NSF-DMS awards 1106862, 1106974 and 1107046), and RDCEP, the University of Chicago Center for Robust Decisionmaking in Climate and Energy Policy (NSF Grant SES-0951576). We thank NASA GISS, NOAA, the Hadley Centre, the IPCC, and IIASA for the use of their publicly available data. We acknowledge the University of Chicago Research Computing Center, whose resources were used in the completion of this work.
%  }\\
    Department of Mathematics and Statistics, Carleton College,\\
    Elisabeth J. Moyer \\
    Department of the Geophysical Sciences, University of Chicago \\
    and \\
    Michael L. Stein\\
    Department of Statistics, University of Chicago}
  \maketitle
} \fi
\if1\blind
{
  \bigskip
  \bigskip
  \bigskip
  \begin{center}
    {\LARGE\bf Estimating trends in the global mean temperature record}
\end{center}
  \medskip
} \fi

\bigskip
\begin{abstract}
Given uncertainties in physical theory and numerical climate simulations, the historical temperature record is often used as a source of empirical information about climate change. Many historical trend analyses appear to deemphasize physical and statistical assumptions: examples include regression models that treat time rather than radiative forcing as the relevant covariate, and time series methods that account for internal variability in nonparametric rather than parametric ways. However, given a limited data record and the presence of internal variability, estimating radiatively forced temperature trends in the historical record necessarily requires some assumptions. Ostensibly empirical methods can also involve an inherent conflict in assumptions: they require data records that are short enough for naive trend models to be  applicable, but long enough for long-timescale internal variability to be accounted for. In the context of global mean temperatures, empirical methods that appear to deemphasize assumptions can therefore produce misleading inferences, because the trend over the  twentieth century is complex and the scale of temporal correlation is long relative to the length of the data record. We illustrate here how a simple but physically motivated trend model can provide better-fitting and more broadly applicable trend estimates and can allow a wider array of questions to be addressed. In particular, the model allows one to distinguish, within a single statistical framework, between uncertainties in the shorter-term versus longer-term response to radiative forcing, with implications not only on historical trends but also on uncertainties in future projections. We also investigate the consequence on inferred uncertainties of the choice of a statistical description of internal variability. While nonparametric methods may seem to avoid making explicit assumptions, we demonstrate how even misspecified parametric statistical methods, if attuned to the important characteristics of internal variability, can result in more accurate uncertainty statements about trends.
\end{abstract}

\noindent%
\vfill

\newpage
%\spacingset{1.45} % DON'T change the spacing!
\section{Introduction}
The physical basis of climate change is understood through a combination of theory, numerical simulations, and analyses of historical data. Climate change is driven by radiative forcing, a change in net radiation (downwelling minus upwelling, often specified at the top of atmosphere) resulting from an imposed perturbation of a climate in equilibrium, for example by increasing the atmospheric concentration of a greenhouse gas. The Earth's response to forcing is complex and not fully understood, in part due to physical uncertainties in important feedbacks such as cloud responses (see the assessment reports of the Intergovernmental Panel on Climate Change (IPCC), e.g.,  \cite{AR5}).

\begin{figure}
\begin{center}
\includegraphics[scale = 0.6]{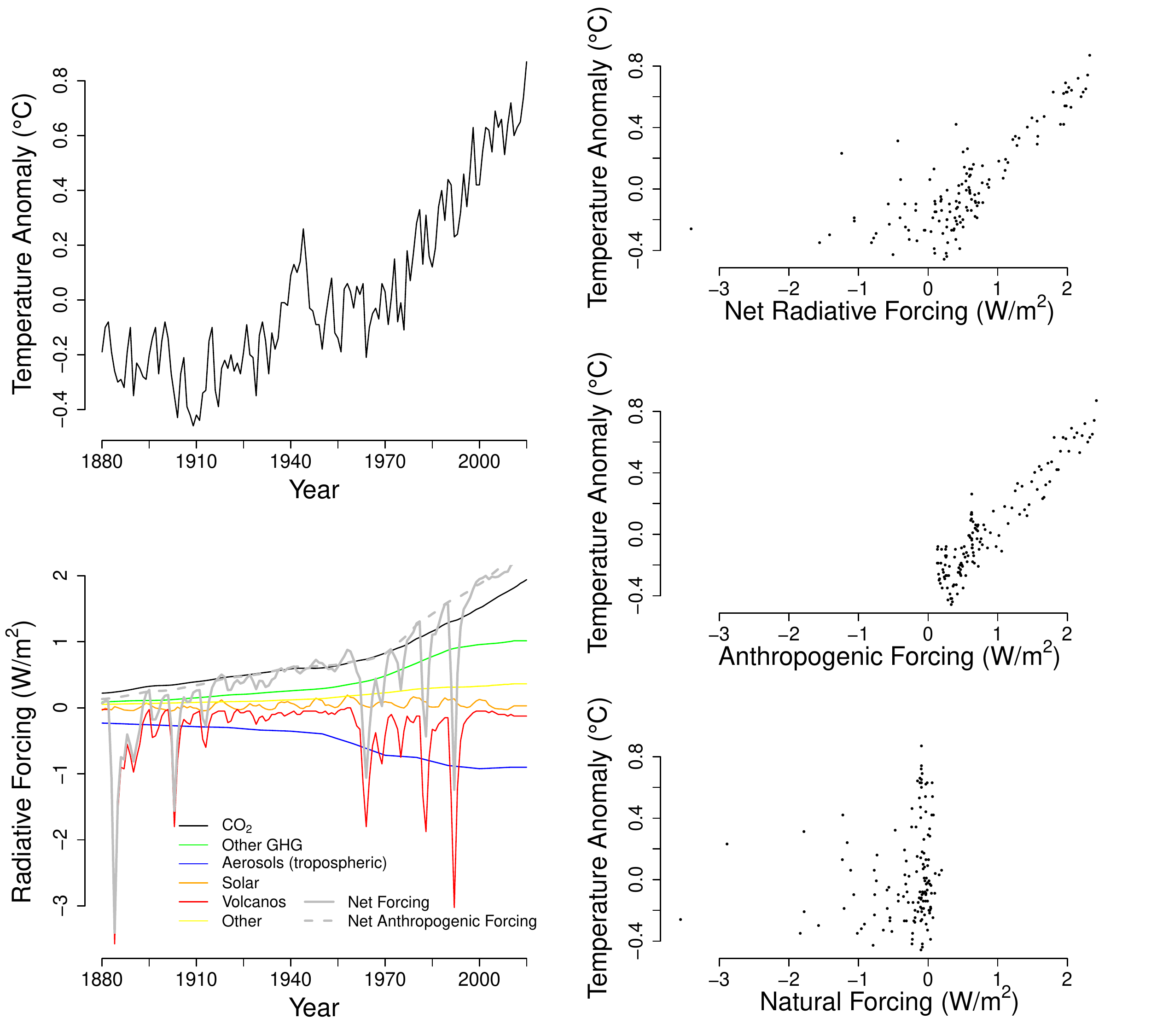}
\end{center}
\caption{Top left, estimates of annually averaged global mean surface temperature anomalies (relative to a base period of 1951-1980) from the years 1880 to 2015. Bottom left, estimates of (top-of-atmosphere) effective radiative forcings from different constituents over this time period (the ``other'' category includes O$_3$, H$_2$O, black carbon, contrails, and land use changes). Right, temperature anomalies vs.\ net radiative forcings (top), vs.\ anthropogenic forcings (middle), and vs.\ natural forcings (bottom). Data sources are described in Section~\ref{sec:data}. Despite the correlation in the plots of temperature vs.\ radiative forcing, temperatures will depend on the full past trajectory of radiative forcings in a potentially complex way, as we discuss in Section~\ref{sec:model}.}
\label{fig:DataPlot}
\end{figure}

Given the physical uncertainties inherent in all climate simulations, the observed temperature record since the late nineteenth century is often used as a source of empirical information about the Earth's systematic response to forcing.  (Figure~\ref{fig:DataPlot} shows one estimate of annually averaged global mean surface temperatures from the past 136 years, along with estimates of radiative forcings from various constituents during that period, with the data sources  described in Section~\ref{sec:data}.) Analysis of the observed temperature record is complicated, however, by the short available record of direct measurements, by uncertainties in the historical radiative forcings themselves, and by the internal temperature variability that exists even in the absence of forcing. Statistical methods are therefore required to quantify the information in the historical record about the response to forcing: given the data, what do we know about how global temperatures have warmed in response to forcing, how much warming can we expect in plausible future forcing scenarios, and how do we expect uncertainties to change as we continue to observe the Earth's temperatures?

It can be helpful to divide approaches to using the observed temperature record to understand aspects of mean climate change into two categories. One common approach involves assuming a physical model of the system. (Here the term ``model'' encompasses anything from a simple energy balance model to a very complicated atmosphere-ocean general circulation model (GCM)). Analysis then may involve estimating statistical parameters in order to best fit the observed record. Estimated parameters have statistical uncertainty because of the finite observational record and the internal variability inherent in the climate system, even were the model a perfect representation of reality. Analyses of observed temperatures using simple or moderately complex physical models include~\cite{wigley1997},~\cite{gregory2002},~\cite{knutti2002},~\cite{forest2002},~\cite{forest},~\cite{gregory},~\cite{padilla},~\cite{aldrin},~\cite{otto},~\cite{rypdal2014},~\cite{masters},~\cite{lewis},~\cite{zeng}, and many others. Analyses using GCMs are summarized in \cite{AR5} Chapter 10 and references therein. 

Another category of approaches involves analyses that are more empirical and appear to deemphasize assumptions about the underlying physics generating the observed temperatures. Many studies use regression models that treat time rather than radiative forcing as the covariate. This practice is often used, for example, to test for significant warming (e.g., \cite{bloomfield}, \cite{smith}, \cite{lovsletten}, and others) or for changes in warming trends (e.g., \cite{foster}, \cite{cahill}, \cite{rajaratnam}, and others); in general, regressions in time are widespread in the literature (see also~\cite{AR5}, especially Chapter 2 Box 2.2, and references therein). 

In both categories, analyses require characterizing internal variability for the purpose of quantifying uncertainty, a task that also involves assumptions. A typical approach is to assume a statistical model for the dependence structure of the noise, such as assuming an autoregressive moving average (ARMA) noise model with a small number of parameters, which is fit to the residuals of whatever trend model is being used. Some authors, however, argue for nonparametric (resampling or subsampling) methods for time series rather than parametric approaches  (e.g., \cite{gluhovsky} and \cite{rajaratnam})\footnote{In this work we consider a parametric method to be any that makes explicit assumptions about the functional form of the probability distribution that the data come from, described with a finite number of statistical parameters. In particular, for the purposes of this work, we consider the class of low-order ARMA models to be a parametric class. We consider a nonparametric method to be one that attempts to make fewer distributional assumptions about the data and does not involve a parametrized statistical model. }. The argument is again that these approaches are advantageous because they are  ostensibly objective and require fewer assumptions.

However, methods that deemphasize assumptions, be they physical or statistical, can be problematic in the climate setting. While regressions in time are simple to apply and do not appear to make explicit assumptions about how temperatures should respond to forcing, these models both limit what can be learned from the data and can result in misleading inferences. Regressions in time are sensitive to arbitrary choices (such as the start and end date of the data analyzed), cannot be expected to apply over even modestly long timeframes, and cannot in general reliably separate forced trends from internal variability. Furthermore, in accounting for internal variability, nonparametric methods for time series often require long data records to work well, and can be seriously uncalibrated in data-limited settings with strong temporal correlation, such as the setting we are discussing.

In the following, we illustrate two primary points. First, we show that targeted parametric mean models that incorporate even limited physical information can provide better fitting, more interpretable, and more illuminating descriptions of the systematic response of interest compared to approaches that deemphasize assumptions. Second, we show that parametric models for residual (i.e., internal) variation can provide for safer and more accurate uncertainty quantifications in this setting than do approaches that deemphasize assumptions, even if the parametric model is misspecified, as long as the parametric modeling is done with particular attention towards the representation of low-frequency internal variability. We believe that the analysis that we present is informative, even if not maximally so, and we attempt to highlight both complications with our analysis as well as important sources of information about global warming that are ignored in our approach. Parts of our analysis share similarities with others listed above, especially with~\cite{rypdal2014} and~\cite{zeng}. In distinction to previous papers, here we are primarily interested in contrasting what can be learned using physically motivated models versus with those that de-emphasize assumptions, as well as in emphasizing the role that accounting for internal variability plays in inferring uncertainties in mean trends.  Our goals are to indicate directions in which statisticians can incorporate explicit modeling to positive effect and to highlight what we view are some of the important sources of uncertainty and information in this problem.

This article is organized as follows. In Section~\ref{sec:data}, we introduce the data sources used in our analysis. In Section~\ref{sec:model}, we provide some background on modeling the historical global mean temperature record and contrast a simple, minimally physically informed model with a more empirical approach. In Section~\ref{sec:UQ}, we highlight insights that can be gained from the more informed approach, with an emphasis on probing different aspects of uncertainty in trends. In Section~\ref{sec:ParaVsNonpara}, through synthetic simulations, we compare the performance of various parametric and nonparametric methods of uncertainty quantification in the presence of temporal correlation in settings similar to that of the historical temperature record. In Section~\ref{sec:discussion}, we give some concluding remarks.

\section{Background and data}
\label{sec:data}
This analysis requires estimates of historical global mean temperatures and radiative forcings. To the extent that we are interested in how temperatures may evolve in the future (and how uncertainty in the response to radiative forcing evolves as more data is observed), we also need radiative forcings associated with a plausible future scenario. 

\textit{Temperature.} We use  the Land-Ocean Temperature Index from the NASA Goddard Institute for Space Studies (GISS) \citep{gisstemp} in our primary analysis. The index combines land and sea surface temperature measurements to estimate annual average global mean surface temperature anomalies (relative to a base period from 1951-1980), extending from the year 1880 to the present (comprising 136 years in total). Any dataset of global mean temperature anomalies represents an estimate of that quantity and is subject to some uncertainty. Sources of uncertainty include the spatial coverage of the network of measurements, interpolation schemes used to estimate temperatures at unobserved locations, methods used to incorporate different sources of data (e.g., land- vs.\ satellite- vs.\ surface buoy- vs.\ ship-based data), and instrumental errors. Sources of uncertainty in the GISS dataset are discussed in ~\cite{gisstemp}. NASA GISS has made some attempt to provide pointwise uncertainty estimates for their data (e.g., Figure 9(a) of~\cite{gisstemp}), but it is important to realize that errors will be correlated in time.  That said, uncertainties in the global mean temperature record are relatively small compared to the changes in temperatures observed over the 20th century.

To evaluate the effects of uncertainty in the temperature record, we repeat a small portion of our analysis using the HadCRUT4 global annual temperature ensemble~\citep{hadcrut4}, designed for that purpose (Section~\ref{subsec:dataUncertainty}). (Some sources of uncertainty are, however, common to both data sources; a brief comparison of these two sources, as well as one from the National Oceanographic and Atmospheric Administration (NOAA), can be found in~\cite{AR5}, Section 2.4.3.)

\textit{Radiative forcing.} The primary driver of climate change during the historical period is changing atmospheric CO$_2$ concentrations; radiative forcing due to these changes scales approximately with the logarithm of the ratio of the CO$_2$ concentration at any given time to its preindustrial level (e.g.,~\cite{arrhenius} and many others). However, variations in other agents also have non-negligible forcing effects that must be taken into account to interpret the historical record (e.g., aerosols from human sources or volcanoes, other greenhouse gases, etc.). In this work, we aggregate the effects of different forcing agents by using their estimated \textit{effective radiative forcing}, the radiative imbalance after rapid atmospheric adjustments. These adjustments are intended to partially compensate for differing efficacies of forcing agents. In practice, effective forcings are often treated as though they may be combined additively. For effective radiative forcings from 1750-2011, we use the estimates in~\cite{AR5} Table AII.1.2 (Figure~\ref{fig:DataPlot} only shows the forcings after 1880, but the full available record is used in our analysis). From 2011-2015, we use the global CO$_2$ concentrations from NOAA and treat radiative forcing from other sources as constant during this period. 

While historical concentrations of CO$_2$ are relatively well known, since CO$_2$ is well-mixed and long-lived, those of other forcing agents, such as tropospheric aerosols, are more difficult to measure because they are spatially heterogenous and short-lived. Uncertainties in forcings associated with tropospheric aerosols are important because aerosol effects can be negatively confounded with greenhouse gas effects (see Figure~\ref{fig:DataPlot}, bottom left). Generally, uncertainties vary by constituent, as does the extent to which the estimates are derived from model output versus observations; see~\cite{AR5} Chapter 8. The focus of this paper is on the information content of the observed temperature record assuming known forcings, but we make a limited attempt to discuss the effect of uncertainty in the forcings (Section~\ref{subsec:dataUncertainty}).

For a plausible future radiative forcing scenario, we use the extended Representative Concentration Pathway scenario 8.5 (RCP8.5)~\citep{rcp85,rcpExtended}, where the change in radiative forcing from the preindustrial level is 8.5 W/m$^2$ by the year 2100 and levels off at around 12 W/m$^2$ in the year 2250. In our simulations, we slightly increase (by about 0.07 W/m$^2$) the radiative forcings from the RCP8.5 scenario in the 21st century to match what we take to be the historical value in 2015, and we assume that natural forcings remain constant after 2015.

\textit{Ocean heat uptake.} The analysis here focuses on information provided solely by the global mean temperature record and assumed known forcings. We do not use additional potential sources of empirical information, including estimates of ocean heat uptake (discussed in, e.g., \cite{forest2002} and \cite{knutti2002}). Many empirical analyses of the historical record do incorporate information about ocean heat content (e.g., ~\cite{gregory2002},~\cite{otto},~\cite{masters}, and ~\cite{lewis}). We compare results with these studies in Section A1.

\section{Modeling trends in the observed global mean temperature record}
\label{sec:model}
Evaluating the systematic response of global mean surface temperatures to forcing is complicated by the long timescales for warming of the Earth system. Because the Earth's climate takes time to equilibrate, the near-term (transient or centennial-scale\footnote{Since the term \textit{transient climate response} has a specific definition in the literature (see the Appendix for a discussion), we use the term \textit{centennial-scale response} to describe the systematic response of temperatures to forcings on the mixing timescale of the mixed layer of the ocean but not of the deep ocean.}) climate response will be less than the long-term (equilibrium or millennial-scale) response. The evaluation is also complicated by the fact that historical radiative forcings are not constant but rather evolve in time (e.g., atmospheric CO$_2$ increases). The physical lags in response imply that the Earth's global mean temperature at any given time depends on the past trajectory of radiative forcings (because the climate does not instantly equilibrate to the present forcing).

A common framework is to decompose observed temperatures into two components: a systematic component changing in response to past forcings and a residual component representing sources of internal variability. That is, for global mean temperatures $T(t)$ at time $t$,
\begin{equation}
T(t) | \{F(t'); t' \leq t\} = f(F(t'); t' \leq t) + \epsilon(t),
\label{eq:trueModel}
\end{equation}
where $f$ is an unknown functional of $\{F(t'); t' \leq t\}$, the collection of past radiative forcings associated with each forcing agent, and $\epsilon(t)$ is a residual process that has mean zero and is correlated in time. Here we emphasize in our notation that the systematic response of interest is the response \textit{conditional on a given forcing trajectory}. The problem, then, is how to estimate the systematic response $f$ using the historical temperatures and forcings. 

\subsection{Regression models in time}
Estimating a model like~\eqref{eq:trueModel} is intractable without additional assumptions. As discussed above, one approach is to resort to physical models. But if instead a more empirical analysis of the observed data is desired, it is common practice to consider a surrogate regression on time itself, as stated in Section 1. The implicit assumption here is that, when viewed as a function of time, the systematic response to historical radiative forcings is approximately linear in time, at least over the considered timeframe:
\begin{equation}
f(F(t'); t' \leq t) \approx \alpha + \beta t, \, \, \, t\in[t_0,t_1]
\label{eq:timeModel}
\end{equation}
where $\alpha$ and $\beta$ are unknown parameters and the trend is considered over the interval $[t_0,t_1]$. The linear time trend approach arguably involves assumptions about the forcing history in addition to the systematic response $f$: that the forcing itself evolves approximately linearly in time (else the approximation would not be appropriate).

The linear time trend model is widely used, and the general sense is that such a model offers a way of testing for statistically significant changes in mean temperature without having to make physical assumptions and without having to believe that the true forced response is linear in time (e.g. \cite{bloomfield} and \cite{lovsletten}). The IPCC, accounting for the apparent appeal of the linear time trend model, writes that it ``is relatively simple, transparent and easily comprehended, and is frequently used in the published research assessed here,'' (\cite{AR5} Chapter 2, Box 2.2), but suggests that linearity in time can at best be viewed as an approximation expected to hold over a relatively short period of time. It is well understood that neither the observed temperature record nor the forcing history appear to evolve linearly over the full range of the data record (Figure~\ref{fig:DataPlot}, left), and most users of the linear time-trend approach confine their analysis to only the past few decades.

While the time trend model may be routine to apply, appear objective, and provide a good fit to the data, its use can be precarious. A proper accounting of uncertainty in mean temperature changes relies on distinguishing internal variability from systematic responses. The time trend model is problematic in this respect. If the chosen time interval is short, it can be difficult to distinguish between trends and sources of internal variability that are correlated over longer timescales than the chosen interval (implicitly recognized in, e.g., \cite{easterling} and \cite{santer} and explicitly discussed in a broader context in \cite{wunsch}). If the chosen interval is long and the systematic trend is actually nonlinear in time, then assuming a linear model in time will shift part of the systematic response to the residual process and can therefore give the impression of excessive internal variability over long timescales (and hence excessive uncertainty in trends).

Because the time trend model cannot be applied over long time intervals for arbitrary forcing scenarios, it also does not have a property that may be considered important for making inferences: that we can learn more about the systematic trend of interest by collecting more observations. There will be only a finite amount of information about the systematic response within the interval $[t_0,t_1]$ (this because sources of internal variability will be positively correlated in time). While this on its own does not invalidate the use of such a model over some narrow time frame, it does mean that what can be learned from the linear time trend model is necessarily limited. More broadly, since the linear time trend model does not map to a physical understanding of the relationship between radiative forcing and global mean temperatures, either during the time interval $[t_0, t_1]$ or extending beyond it, the questions that can be asked with this model are narrow. 

Some argue that many of these problems may be overcome by using a model that is nonlinear in time, such as a spline or other nonparametric regression method. (The IPCC, for example, appears to view nonparametric extensions as more generically appropriate than the linear model.) Nonparametric regressions in time will appear to provide an even better fit to the data than the linear trend model, but many of the above arguments carry over to this setting. Such models have limited interpretational value or ability to capture systematic (non-internal) trends, since they cannot generically be expected to distinguish between the systematic trends of interest and other, internal sources of long-timescale variation in the data. Collectively, these arguments suggest that it is advisable to seek better motivated models if one is interested in understanding the systematic response of global temperatures to forcing.

\subsection{A simple, physically-based model for the centennial-scale response to forcing}
A typical approach is to use more complex models, including full GCMs, to explain the systematic response of interest. (Model output is also used in concert with observations in the context of ``detection and attribution'' studies; see, e.g., Chapter 10 of~\cite{AR5}.) Some may object to this approach, however, out of a worry that the climate model has already been tuned to match the observed historical temperature trend or is otherwise conditioned on past temperature observations~\citep{knutti,huybers,mauritsen}. There is therefore value in a compromise approach between the linear time trend model and very complex numerical simulations. In this work, we discuss a statistical model that is easy to apply but that encodes some physical intuition for the problem that makes the model interpretable and hopefully applicable over longer time periods. The goal is to show that even simple models incorporating limited physical information can provide more insight about temperature trends and their uncertainties given the observed data than can regression models in time.

A commonly used, very simplified physical model for the response to an instantaneous change in radiative forcing is that temperatures approach their new equilibrium in exponential decay. That is, writing $F_{\text{inst}}(t)$ for a step function that changes at time $t=0$,
\begin{equation}
\label{eq:instResponse}
f(F_{\text{inst}}(t');t'<t)  \approx \mu_0 + \lambda(1 - \rho^t)\mathbf{1}\{t\ge0\},
\end{equation}
where $\lambda$ is the change in equilibrium temperature, $\mu_0$ is the mean temperature in the baseline state, and $\rho$ controls the rate at which the changes in temperatures approach $\lambda$, taking values between zero (instantaneous response time) and one (infinite response time). Equation~\eqref{eq:instResponse} represents the response function for a linear model of temperature change, so is a natural first approximation for the evolution of temperature in the case of small perturbations from steady state (e.g., \cite{mackay}). In particular, equation~\eqref{eq:instResponse} can be interpreted as the solution to a simple energy balance model that makes two assumptions: first, that the equilibrium temperature change is linear in the forcing (i.e., that the \textit{linear forcing feedback model} holds), and second, that the rate of warming is approximately proportional to the heat uptake. The model is overly simplified because the Earth shows responses at multiple timescales (e.g., \cite{held,olivie,geoffroy} and others) but can provide a reasonable approximation for the response on timescales shorter than those associated with full equilibration. In any case, when convolved with a time-varying forcing trajectory, the resulting model for the systematic response is an infinite distributed lag model in the forcing trajectory with weights decaying exponentially (e.g., \cite{caldeira,castruccio}). Models based off of~\eqref{eq:instResponse} have previously been considered for analyses of the observed global mean temperature record (e.g.,~\cite{rypdal2014} and \cite{zeng}).

We also use a model based off of~\eqref{eq:instResponse} for the systematic temperature response in the observed data:
\begin{equation}
f(F(t'); t' \leq t) \approx \mu_0 + \lambda_A h\left(\rho_A,\frac{F_A(t')}{F_{2\times}}; t'\leq t \right) + \lambda_N  h\left(\rho_N,\frac{F_N(t')}{F_{2\times}}; t'\leq t \right) ,
\label{eq:emulator}
\end{equation}
where 
\[
h(\rho, x(t'); t'\leq t) = (1-\rho) \sum_{k=0}^{\infty} \rho^k x(t-k).
\]
In model~\eqref{eq:emulator}, $\lambda_A$ and $\lambda_N$ represent ``sensitivities'' to anthropogenic and natural forcings, $F_A$ and $F_N$, respectively, and have units of degrees Celsius temperature change per forcing change $F_{2\times}$ (the forcing associated with a doubling of atmospheric CO$_2$, approximately 3.7 W/m$^2$). The parameter $\lambda_A$ is similar to the \textit{equilibrium climate sensitivity}\footnote{The  change in mean temperature associated with a doubling of CO$_2$ concentration, after a sufficiently long time that the climate has reached a new equilibrium.}, but will be estimated as somewhat lower than that quantity, in part because the proposed model contains only a single timescale of response to anthropogenic forcing (see the Appendix). \cite{caldeira} and \cite{castruccio} used multiple timescales in modeling longer series from GCM output, but we cannot distinguish these with only 136 years of data and given a smooth past trajectory of anthropogenic forcings. Response timescales to anthropogenic and natural forcings are set by the parameters $\rho_A$ and $\rho_N$ (taking values between zero and one). Model~\eqref{eq:emulator} should approximate temperature trends reasonably well up to centennial timescales, but not at the millennial timescales at which the deep ocean mixes.

Our approach differs from those in~\cite{rypdal2014} and \cite{zeng} in some important ways. \cite{zeng} assume that net radiative forcing is a constant multiple of CO$_2$ forcing, treat the $\rho$ parameter as known, and do not account for uncertainty due to internal variations that are correlated in time. \cite{rypdal2014} suggest replacing the standard exponential decay response used in model~\eqref{eq:emulator} with a power-law decay response. The basis for this suggestion is an implicit assumption that we do not make: that the same simplified model for the climate's systematic response to radiative forcing should be used to model residual, internal variability as a response to white noise forcing. More details on our uncertainty quantification for model~\eqref{eq:emulator}, including our model for internal variability, are given in below in Section~\ref{sec:UQ}.

In building model~\eqref{eq:emulator}, we chose to separate natural and anthropogenic forcings because they seem not strictly comparable (Figure~\ref{fig:DataPlot}, right). First, there may be some evidence that aerosol forcing from volcanic eruptions is less efficacious than CO$_2$ forcing~\citep{sokolov,hansen,marvel}. While we use estimates of effective radiative forcing, which should compensate for efficacy, these estimates do not include an adjustment for the volcanic forcing. Moreover, the timescales of response associated with these forcings may also be different, possibly because ocean heat content responds differently to sudden and/or negative changes in forcing (as produced by volcanic eruptions) compared to more gradual and/or positive changes (as in continued anthropogenic emissions of CO$_2$) (e.g., \cite{gregory, padilla}). We combine solar and volcanic forcings out of convenience; the solar forcings do change more rapidly than the anthropogenic forcings, and in any case the magnitude of the changes in solar forcing is small.

\begin{figure}
\begin{center}
\includegraphics[scale = 0.6]{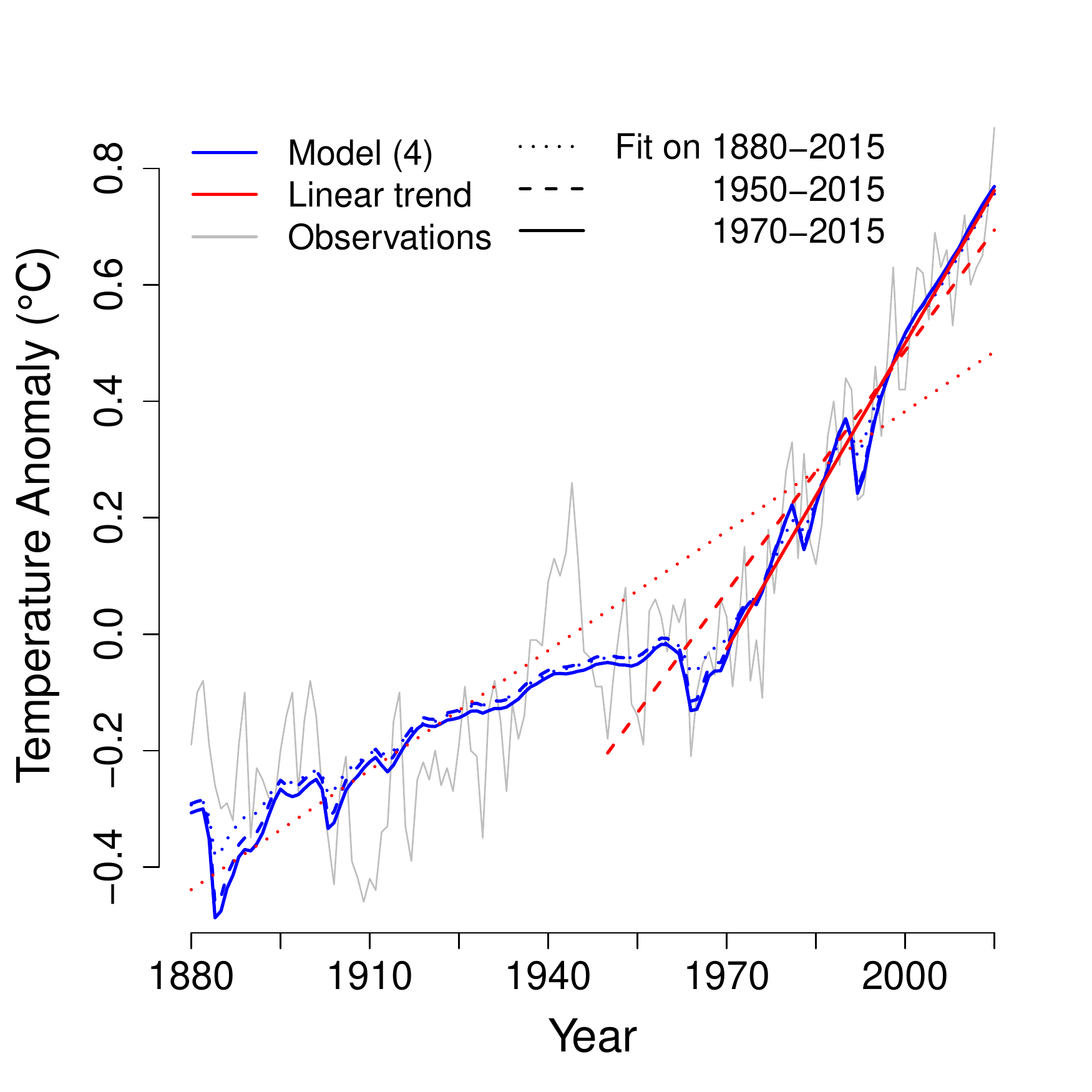}
\end{center}
\caption{Comparison of the fitted values for model~\eqref{eq:emulator} and the linear time trend model fitted to the global mean temperature record over different timeframes. (The fitted model~\eqref{eq:emulator} trends are extended back to 1880 regardless of the timeframe used to fit the model.) The linear model appears in agreement with model~\eqref{eq:emulator} roughly after 1970, but not before. By contrast, model~\eqref{eq:emulator} produces fairly stable estimates of the mean response during the 20th century, although we note that the apparent fit to the data may be slightly poorer in the earliest part of the record.}
\label{fig:LinearVsEm}
\end{figure}

To illustrate the use of model~\eqref{eq:emulator}, we fit it to different segments of the observed global mean surface temperature record and we compare to the linear fits estimated over the same timeframes. Figure~\ref{fig:LinearVsEm} shows the resulting fitted models using the data from 1970-2015, 1950-2015, and 1880-2015. The estimated trend from model~\eqref{eq:emulator} is relatively insensitive to the timeframe used. The estimated linear time trends, on the other hand, differ markedly using different timeframes, and agree with model~\eqref{eq:emulator} only after around 1970, during the time period over which net radiative forcing was evolving approximately linearly in time (see again Figure~\ref{fig:DataPlot}). The sensitivity of the inferred linear trend in global mean temperature to the starting date has been previously discussed (e.g.,~\cite{liebmann}).  These results suggest both that model~\eqref{eq:emulator} does indeed capture important aspects of the underlying physical processes driving temperature trends and that it therefore may be used to answer more interesting questions than can the linear time trend model.

\section{Trend and uncertainty: what can we learn from applying our simple model to data?}
\label{sec:UQ}
In this section, we illustrate what can be learned by applying the simple model~\eqref{eq:emulator} to observed temperatures. To do this, we must introduce an additional model to capture internal variability ($\epsilon(t)$ in the assumed true model~\eqref{eq:trueModel}). We then use our full model to infer the parameters in model~\eqref{eq:emulator}, to evaluate their uncertainties given the data, and to explore the implications for understanding temperature trends.

\begin{figure}
\begin{center}
\includegraphics[scale = 0.5]{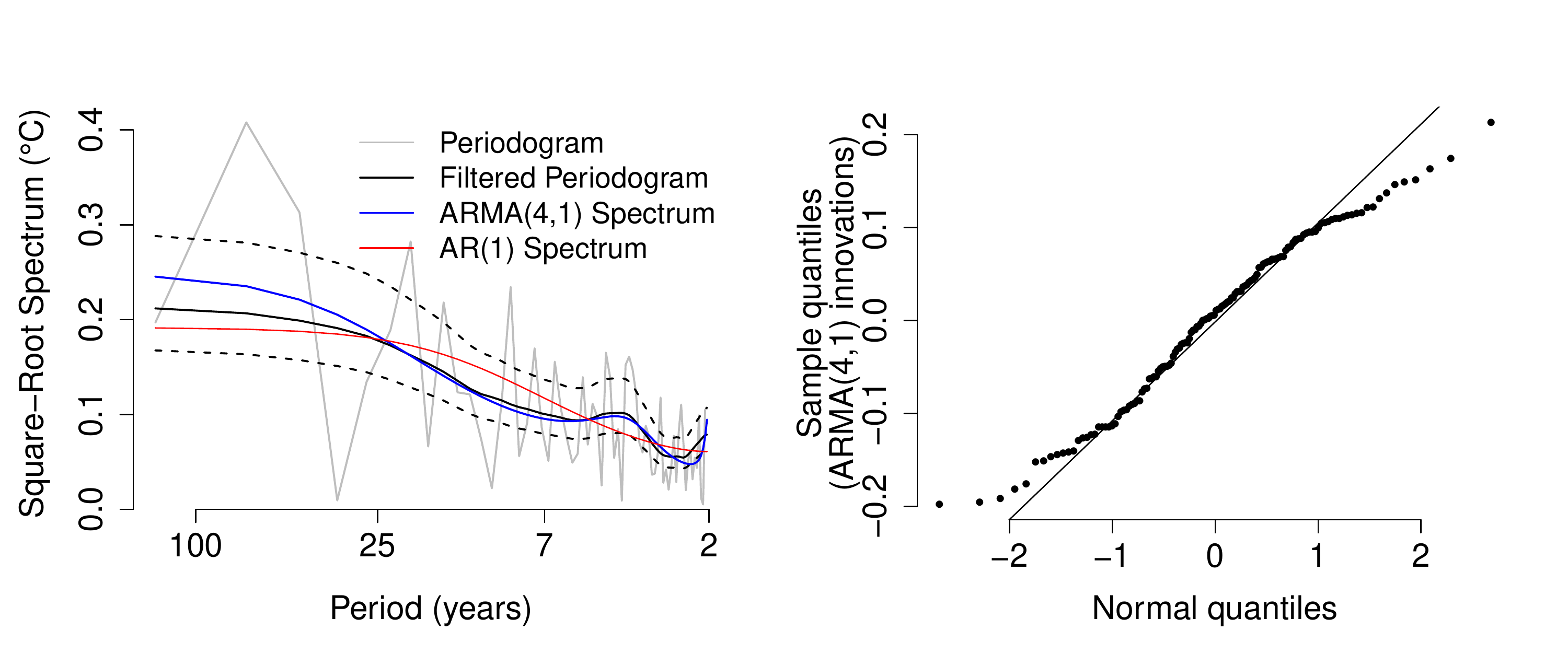}
\end{center}
\caption{Left, the raw periodogram of the residuals from model~\eqref{eq:emulator}, a filtered periodogram (twice applying a moving average of width 5), and the power spectra associated with fitted ARMA(4,1) and AR(1) models. The dashed lines are at $\pm$2 standard errors associated with the filtered periodogram. The ARMA(4,1) model appears to more realistically represent low-frequency variation in the residuals, which is crucial for inferences about trends. Right, normal quantile-quantile plot for the ARMA(4,1) sample innovations. There is some evidence that the innovations are light-tailed compared to the normal distribution.}
\label{fig:resSpectrum}
\end{figure}

To diagnose features of internal variability, spectral analysis is an intuitive framework, since the frequency properties of internal variability are tied to uncertainties in trends: uncertainty in smooth trends is more strongly  affected by low-frequency than high-frequency internal variability. Figure~\ref{fig:resSpectrum}, left, shows the raw periodogram associated with the residuals from model~\eqref{eq:emulator} and a smoothed estimate of the spectrum. We also show the spectra associated with fitting the residuals to two models for the internal variability. Both are standard autoregressive-moving average (ARMA) models (Appendix Section A4 includes a definition), allowing for dependence between the noise at each time step with the noise for past values. The two models we compare are an ARMA(4,1) and an AR(1) model, which were chosen according to common information criteria used for selecting time series models: respectively, the small-sample corrected Akaike information criterion (AICc)~\citep{hurvich} and the Bayesian information criterion (BIC)~\citep{schwarz} (See Table A2 for the coefficient estimates, innovation standard deviations, and AICc and BIC associated with these two models.) Both models appear to fit the data reasonably well; the ARMA(4,1) model arguably overfits at the higher frequencies, but the AR(1) model may be underestimating variability at the lowest frequencies.  Since low frequency variability is most important to uncertainties in smooth trends, we adopt the more conservative choice of using the ARMA(4,1) model. Figure~\ref{fig:resSpectrum}, right, shows the normal quantile-quantile (Q-Q) plot for the sample innovations from this model;  there is evidence that the innovations are somewhat more light-tailed than Gaussian, so standard errors based on a Gaussian assumption should not be overoptimistic.

\begin{figure}
\begin{center}
\includegraphics[scale = 0.4]{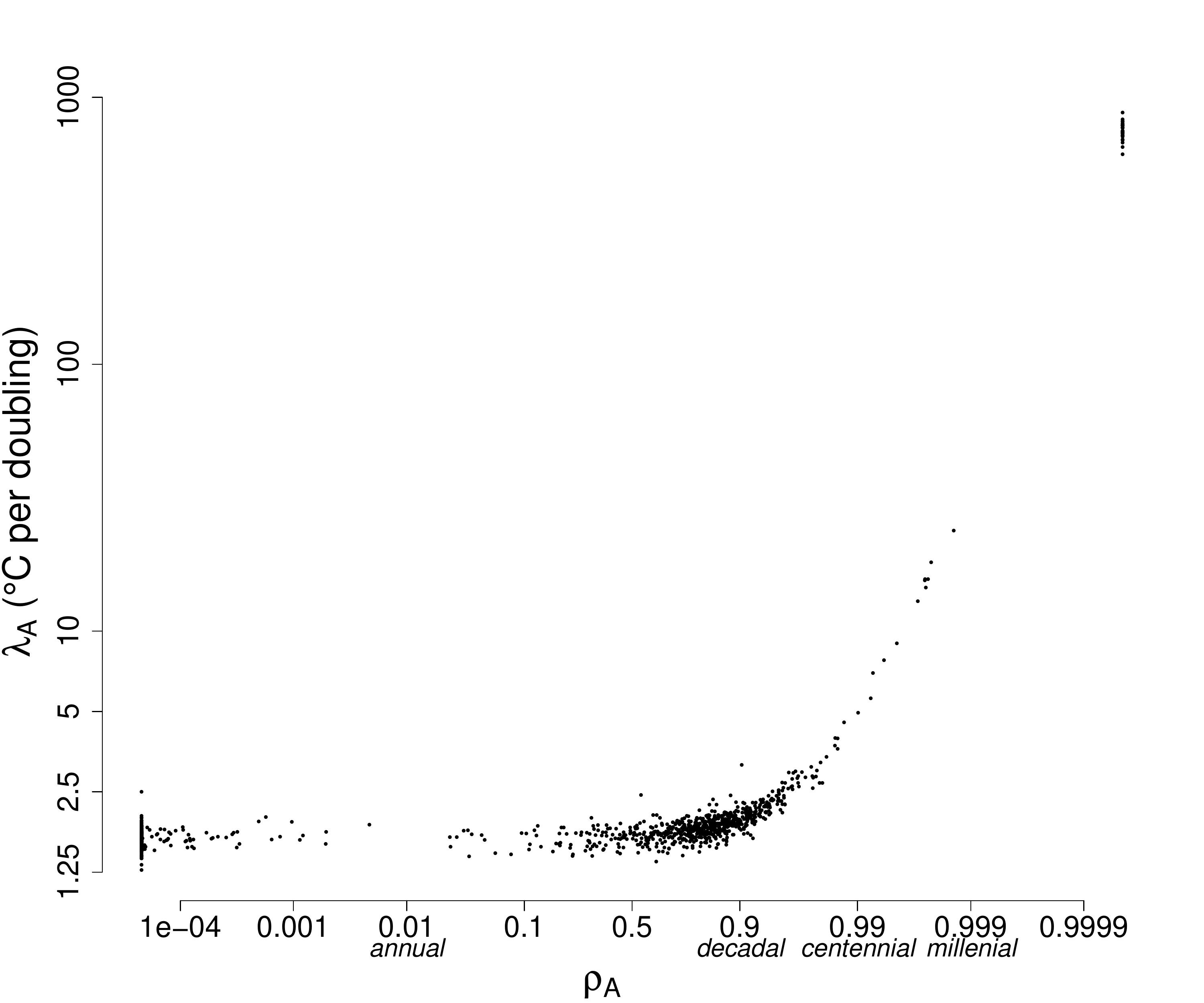}
\end{center}
\caption{Distribution of the parametric bootstrap estimates of $\lambda_A$ and $\rho_A$ from model~\eqref{eq:emulator}. It is difficult to distinguish between the rate of response and the sensitivity using only global mean temperatures from recent history.}
\label{fig:bootDist}
\end{figure}

Using the fully parametric model, combining~\eqref{eq:emulator} and a Gaussian ARMA(4,1) noise model, we proceed with uncertainty quantification through a parametric bootstrap\footnote{A parametric bootstrap involves generating repeated, synthetic simulations under the fitted statistical model and then refitting the model to each simulated time series to obtain new estimates of model parameters. The distribution of those estimates then gives a measure of the uncertainty in the original estimates.}. When applied to our model of the historical temperature record, the parametric bootstrap distribution shows, unsurprisingly, that in a relatively short time series and given a smooth past trajectory of forcings, it is difficult to distinguish between a climate with both a high sensitivity (large value of $\lambda$) and slow response (large value of $\rho$) versus one with a lower sensitivity (smaller value of $\lambda$) but a faster response (smaller value of $\rho$).  The estimates $\hat{\lambda}_A$ and $\hat{\rho}_A$ are therefore strongly dependent, with $\hat{\lambda}_A$ increasing explosively as $\hat{\rho}_A\to1$ (Figure~\ref{fig:bootDist} shows their bivariate parametric bootstrap distribution). The strong nonlinear relationship between these two parameters, and the high degree of skewness in the marginal distribution of $\hat{\lambda}_A$, are the reasons that we rely on bootstrapping for uncertainty quantification, as the typical appeals to asymptotic normality are not viable in this setting.

In the following, we will represent uncertainties using the simple bootstrap percentile method.  The percentile method is subject to criticism (e.g., \cite{hall}). We have found, however, that adjusting the percentile method using a nested bootstrap results in narrower confidence limits. For this reason, we believe that the raw percentile-based intervals may be conservative in this setting, so we choose to report the apparently conservative intervals. Our point estimates are based on a two-step procedure wherein model~\eqref{eq:emulator} is estimated via least squares and the ARMA(4,1) model is estimated via maximum likelihood on the residuals. While this procedure may be somewhat suboptimal compared to jointly estimating the mean and covariance structure, the two-step procedure is substantially faster, which is important for carrying out the nested bootstrap.

\subsection{Uncertainties in the sensitivity parameters}
When using the full 1880-2015 global mean surface temperature record, the point estimate for the centennial-scale sensitivity to anthropogenic forcing is $\hat{\lambda}_A=1.8^\circ$C per doubling, with $\hat{\rho}_A=0.80$ (which implies mixing on decadal timescales). The estimated sensitivity to natural forcing is much smaller, $\hat{\lambda}_N = 0.21^\circ$C per doubling with $\hat{\rho}_N = 0.58$. In this section, we discuss uncertainties in these estimates.

Using our statistical model, the historical data appear to provide a lower bound for $\lambda_A$ (assuming for now that the forcings are known) but cannot rule out extremely large and implausible values on the order of tens or hundreds of degrees per doubling (the 2.5-97.5th bootstrap percentile interval is 1.5 to 690$^\circ$C per doubling). These very large values are not supported by evidence from the paleoclimate record (e.g.,~\cite{AR5} Chapter 5) and the approximation of the linear forcing feedback model, on which~\eqref{eq:emulator} implicitly relies, breaks down under high sensitivity~\citep{blochjohnson}. The large upper bound should therefore be interpreted only as a statement about the information in the historical global mean temperature data under the strict assumption that model~\eqref{eq:emulator} holds exactly.  For the sensitivity to natural forcings, the data cannot rule out $\lambda_N=0$ (the 2.5-97.5th bootstrap percentile interval is -1.1 to 4.1$^\circ$C per doubling). Table~\ref{tab:SensitivityCIs} gives some intervals at different percentiles for the parameters $\lambda_A$ and $\lambda_N$. (The parameters $\rho_A$ and $\rho_N$ are essentially unconstrained by the data.)

\begin{table}[t]
\caption{Parametric bootstrap percentile intervals for the sensitivities in model~\eqref{eq:emulator}, $\lambda_A$ and $\lambda_N$, to anthropogenic and natural sources, respectively (in units $^\circ$C per doubling). The data appear to provide a lower bound for $\lambda_A$, but cannot rule out even implausibly large values; the very large values are associated with slow responses (see Figure~\ref{fig:bootDist}). The data cannot rule out $\lambda_N=0$, likely because there are few prominent volcanic eruptions in the historical record analyzed and the response to volcanic aerosols may be small.}
\label{tab:SensitivityCIs}
\begin{center}
\begin{tabular}{l | l l | l l}
\hline\hline
& \multicolumn{2}{c|}{$\hat{\lambda}_A=1.8$} &  \multicolumn{2}{c}{$\hat{\lambda}_N=0.80$}  \\
Percentile interval & Lower & Upper & Lower & Upper \\
\hline
5-95 (90\%) & 1.5   & \hphantom{0}3.0 & -0.15 & 1.5 \\
2.5-97.5 (95\%) &  1.5    & 690  & -1.1 & 4.1 \\
0.5-99.5 (99\%) & 1.4 & 790 & -13 & 490 \\
\hline
\end{tabular}
\end{center}
\end{table}

The IPCC's own 66\% ``likely'' interval for equilibrium sensitivity is $1.5^\circ$C to $4.5^\circ$C per doubling, which subjectively combines estimates from various sources using multiple lines of evidence including from ensembles of models with different physics, and accounts for other sources of uncertainty that we have so far ignored, such as uncertainty in the forcings themselves (see~\cite{AR5} 10.8.2 and Box 12.2). The bulk of the distribution of our estimate is somewhat narrower than the IPCC's, likely because so far we have not accounted for uncertainty in radiative forcings; however, the IPCC rules out the very large and unphysical values in the right tail of our intervals. We here stress again that since we are estimating the centennial-scale response, the estimates of the sensitivity that we provide will tend to be lower than the equilibrium sensitivity estimated in the IPCC's interval (see the Appendix, which also compares our estimates to other observationally based estimates of a sensitivity parameter). Individual estimates of the equilibrium climate sensitivity and associated uncertainty in the literature are discussed in \cite{AR5} Section 10.8.2 (see also again the Appendix).

The main source of uncertainty in the upper bound for $\lambda_A$ is due to uncertainty in the ``equilibration time'' of the climate associated with smoothly increasing anthropogenic forcing, controlled by $\rho_A$. If we restrict $\rho_A$ to, say, centennial scales or smaller, then the uncertainty is substantially decreased (see again Figure~\ref{fig:bootDist}). One may argue through other lines of evidence and reasoning that extremely large values of $\rho_A$ and $\lambda_A$ are implausible, but the statistical model is being used to quantify the information content of the historical temperature record. The fact that additional sources of information are needed to exclude unphysical values suggests that even our minimally informed model may be overly empirical for some purposes. The inability of the data to rule out $\lambda_N=0$ is, on the other hand, probably due to the fact that there are few prominent volcanic eruptions in the historical record analyzed and that the response to volcanic aerosols may be small for the reasons discussed above.

\subsection{Uncertainties in near- and long-term trends}
\begin{figure}
\begin{center}
\includegraphics[scale = 0.50]{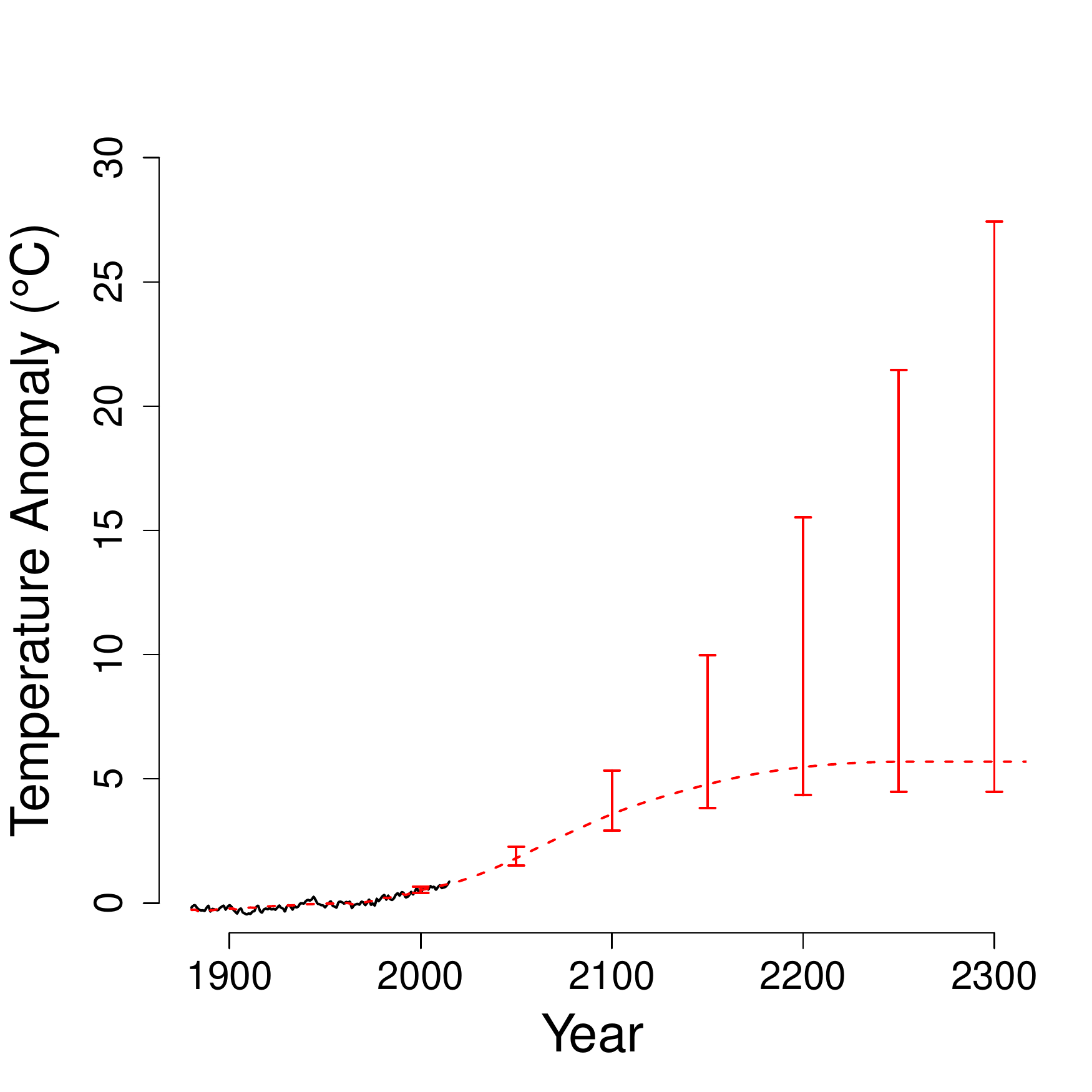}
\end{center}
\caption{Projected mean temperature anomalies, and their uncertainties, under the RCP8.5 scenario, based on estimates from model~\eqref{eq:emulator} and assuming ARMA(4,1) noise. The black curve shows the observed temperatures. Intervals are pointwise (2.5-97.5)-percentile intervals. Radiative forcing stabilizes in the year 2250, but mean temperatures and especially their uncertainties continue to increase. While uncertainties in the long-term response are quite large, due largely to the inability to rule out implausible values of $\lambda_A$, the historical and near-term response is much more certain.}
\label{fig:projPlot}
\end{figure}

The uncertainties in the sensitivity and rate of response parameters imply greater uncertainties in projected longer-term future trends in global mean temperature than in the historical and near-term projected trends. To illustrate this, we examine the implied future trends under the hypothetical (extended) RCP8.5 scenario, in which radiative forcing increases and then stabilizes in the year 2150. We simulate new time series using our estimates of model~\eqref{eq:emulator} and the ARMA(4,1) noise model, given radiative forcings from this scenario. The projected trend and associated pointwise uncertainties are shown in Figure~\ref{fig:projPlot}.

Projected mean temperatures, and especially their associated uncertainties, continue to increase even after stabilization of forcing. This is a consequence of the joint uncertainty in $\lambda_A$ and $\rho_A$, and in particular of the inability to rule out implausible values of these parameters. If the goal, then, is to provide a long-term projection of mean temperatures given only the historical temperature record, these estimates will unsurprisingly be quite uncertain (even assuming known past and future forcings).

On the other hand, trends in the historical and near-term response are much more certain. The observations strongly suggest that mean temperatures increased in the 20th century: for example, the (2.5-97.5)-percentile interval for the mean response in the year 2000 (expressed compared to the 1951-1980 average) is well above zero at (0.4,0.6)$^\circ$C. (The 30 year average of the data around the year 2000 gives an anomaly of 0.5$^\circ$C, in line with our estimate using model~\eqref{eq:emulator}.) These kinds of distinctions between the uncertainty in the near- and longer-term mean responses are not easily made using a time trend model.

\subsection{Decreasing uncertainty in the sensitivity parameter as more data is observed}
We have shown that the short historical temperature record alone produces fairly uncertain estimates of the sensitivity parameter, $\lambda_A$ in model~\eqref{eq:emulator} (Figure~\ref{fig:bootDist}, Table~\ref{tab:SensitivityCIs}), and therefore of longer-term temperature trends (Figure~\ref{fig:projPlot}). We now examine how these uncertainties decrease as the temperature record increases (as in, e.g., ~\cite{kelly, ringLearning, padilla, urban, myhre} and others). To do this, we artificially extend the temperature record by generating new synthetic time series using the mean and noise models estimated from the historical data and forcings from the same RCP8.5 scenario described above. We then reestimate model~\eqref{eq:emulator} for each synthetic series, using successively longer synthetic datasets. The results suggest that the data will not constrain the upper bound on the sensitivity parameter until another $\sim$50 years, by which time (under our estimated model and the RCP8.5 scenario) temperatures will have already risen by about 3$^\circ$C from the preindustrial climate. A summary of the evolution of uncertainties is given in Table~\ref{tab:SensByYearCIs}. 

\begin{table}[t]
\caption{Evaluation of how uncertainty in the sensitivity parameter, $\lambda_A$, decreases as more data is observed, from simulations under the RCP8.5 scenario of future radiative forcing with an assumed value of $\lambda_A = 1.8^\circ$C per doubling. The middle column shows the (2.5-97.5)th-percentile intervals for $\lambda_A$ from simulations under the fitted model~\eqref{eq:emulator} and ARMA(4,1) noise. The rightmost column shows the increase in mean temperatures from the preindustrial climate under the fitted model at the year in question. Uncertainties in the upper bound of $\lambda_A$ decrease relatively slowly as more data is observed.}
\label{tab:SensByYearCIs}
\begin{center}
\begin{tabular}{l | l l }
\hline\hline
\multirow{2}{*}{End year} & (2.5-97.5)th percentile  & Change in mean \\
& ($^\circ$C per doubling) & from preindustrial ($^\circ$C) \\
\hline
2025 & (1.5, 16) & 1.3 \\
2050 & (1.6, 11)   & 2.2\\
2075 & (1.6, 2.8) & 3.1 \\
2100 & (1.7, 2.0) & 3.9 \\
\hline
\end{tabular}
\end{center}
\end{table}

These estimates could be more strongly constrained by using additional physical information. As discussed previously, the very high sensitivity estimates in the bootstrap distribution are cases where the estimated response time is unphysically long. Without external information about this timescale, however, long data records are required to rule out the large values of $\rho_A$ and $\lambda_A$ that the model entertains.

\subsection{Is there evidence of long memory internal variability in global mean temperatures?}
One of the complicating factors in estimating trends in climate time series is the question of whether global mean temperatures exhibit \textit{long memory}. Long memory processes have power spectra that behave like $(2\sin(\omega/2))^{-2d}$ as the frequency $\omega\to0$, with $d>0$ (i.e., infinite power at the frequency zero). When $d<1$ the process has finite total variance, as would be expected for a variable like global mean temperature. For more details on long memory processes, see~\cite{beran2013}. By contrast, short-memory processes (like the ARMA(4,1) model we assume), have finite power at the origin. Many authors have suggested that internal temperature variability is well-modeled by processes with long memory but finite variance (e.g., ~\cite{frankignoul},~\cite{bloomfield},~\cite{smith},~\cite{gil},~\cite{lennartz},~\cite{fraedrich},~\cite{rypdal2014},~\cite{lovsletten}, and many others). Some authors have moreover claimed that global mean temperatures are well-modeled by a random walk (e.g., \cite{gordon}), as has at least one standard time series textbook~\citep{shumway}, which would imply that global mean temperatures do not have a finite variance over time.  In either case, if the Earth's temperatures exhibited long memory, it would be more difficult to estimate trends than in the short memory case, since low-frequency variability would then be more difficult to distinguish from trends. 

\begin{figure}
\begin{center}
\includegraphics[scale = 0.6]{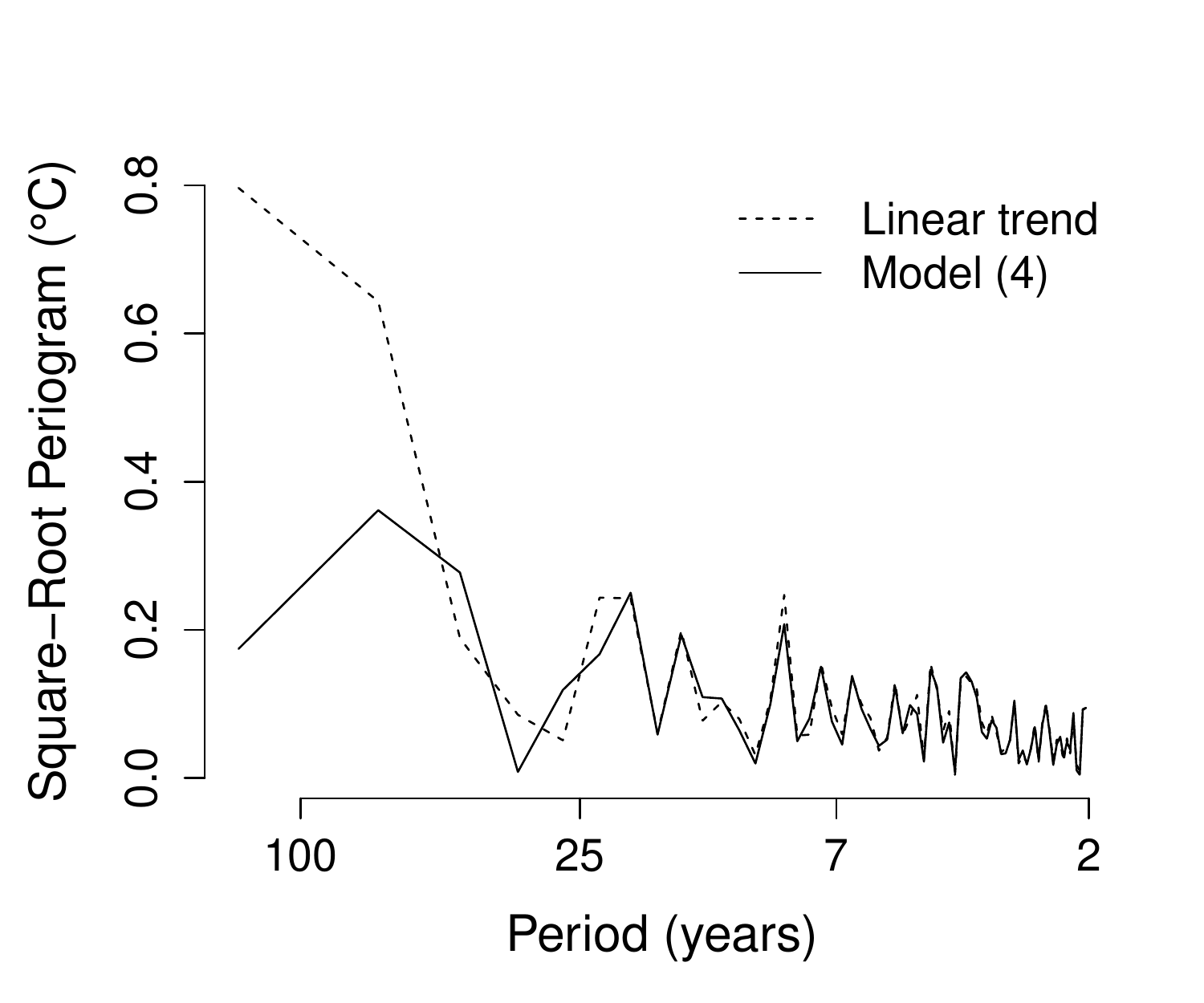}
\end{center}
\caption{Raw periodograms of residuals from models~\eqref{eq:timeModel} (dashed) and~\eqref{eq:emulator} (solid) fit to the full data record. There is substantially more low-frequency variability in the residuals from the linear trend model than in the residuals from model~\eqref{eq:emulator}. Misspecified mean models will give a misleading impression of low-frequency variability, and therefore misleading uncertainties associated with the mean trend.}
\label{fig:LinearVsEmulatorSpectrum}
\end{figure}

The evidence for long memory, however, strongly depends on the assumed trend model. Many of the aforementioned authors draw their conclusions by assuming a linear time trend model and applying that model to the temperature record on durations of decades to over a century. (One notable exception is \cite{rypdal2014}, but they only compare long memory noise models to AR(1) models.) As discussed previously, a linear trend model applied to a time series with a nonlinear trend will imply excessive low-frequency noise. Figure~\ref{fig:LinearVsEmulatorSpectrum} shows the periodograms of the residual global mean temperatures after removing either a linear time trend or a trend of the form of~\eqref{eq:emulator}. While the high-frequency behavior of the residuals is not much affected by the choice of trend model, the low-frequency behavior is very much affected. Apparent low-frequency variability is made more severe by assuming that mean temperatures increase linearly in time.

The question of long memory cannot be definitively settled using a dataset of only 136 observations, and other analyses make use of longer climate model runs or the paleoclimate record (e.g.,~\cite{mann}). Nevertheless, it should be clear that the linear time trend model is especially problematic for this purpose. In general, regression models in time, linear or otherwise, have a danger either of mistaking systematic trend for apparent low frequency variability (as just described), or of mistaking low-frequency variability for systematic trend (as would occur, for example, when using a nonparametric regression with too small of a smoothing bandwidth). Either can lead to misstated uncertainties, and therefore can be problematic even if the claim is that the trend model is only being used to test for significant warming and that the model is not believed to be true.

\subsection{Implications of uncertain inputs: radiative forcing and temperatures}
\label{subsec:dataUncertainty}
The analysis thus far has assumed that both radiative forcings and temperatures are known exactly, but uncertainty in the sensitivity and in trends also propagates from uncertainty in these quantities. We therefore discuss at least roughly the potential implications of imperfect knowledge of these inputs.

Of the two factors, uncertainty in radiative forcings, particularly from aerosols, is more consequential, especially for the inferred lower bound of the sensitivity parameter from model~\eqref{eq:emulator}. The importance of radiative forcing uncertainty for uncertainties in climate sensitivity has been widely noted (e.g., ~\cite{gregory,padilla, otto, masters, lewis, myhre}). The trajectory of net effective radiative forcing from anthropogenic sources is poorly known; the IPCC states that the difference in net effective radiative forcing from anthropogenic sources between the years 2011 and 1750 was about 2.29 W/m$^2$ with a 95\% confidence interval of 1.13 to 3.33 W/m$^2$. We explore the implications of this uncertainty by simply scaling the entire trajectory of net radiative forcings from 1750 to the present such that the uncertainty in 2011 is as stated. Adjusting aerosol forcings to the high or low ends, respectively, produces sensitivity estimates from~\eqref{eq:emulator} varying by over a factor of two, from 1.2 to 3.7$^\circ$C per doubling (vs.\ the original estimate 1.8$^\circ$C). The aerosol uncertainty appears important: the value 1.2$^\circ$C per doubling is lower than all of the bootstrap values of $\lambda_A$ generated assuming known forcings (Figure~\ref{fig:bootDist} and Table~\ref{tab:SensitivityCIs}).

Uncertainties in the global mean surface temperature record are comparatively less important. To partially address this issue, we re-estimate model~\eqref{eq:emulator} using each of the 100 ensemble members of the HadCRUT4 global mean temperature ensemble. The point estimates of $\lambda_A$ range from 1.5 to 2.1$^\circ$C per doubling, with estimates of $\rho_A$ ranging from 0.79 to 0.90. The point estimates of $\lambda_N$ range from 0.71 to 20$^\circ$C degrees per doubling, with estimates of $\rho_N$ ranging from 0.86 to 0.996. The additional uncertainty induced by observational uncertainty in the temperature record is smaller than either the uncertainty induced by internal temperature variability or the uncertainty in radiative forcings.

\subsection{Bayesian methods}

Some of the uncertainties discussed so far could be addressed in a Bayesian framework (as in, e.g., \cite{forest2002}, \cite{forest}, \cite{padilla}, and \cite{aldrin}) although we do not pursue that approach here. For example, the effects of uncertainties in temperatures and radiative forcings could be modeled hierarchically, and outside information could be used to constrain long-timescale responses not informed by the data. We have instead chosen here to focus on the information contained in global mean temperature observations about trends to illustrate some important sources of uncertainty in empirical analyses even assuming known inputs.

\section{Parametric vs.\ nonparametric uncertainty quantification}
\label{sec:ParaVsNonpara}
In Sections~\ref{sec:model} and~\ref{sec:UQ}, we showed that analyses of the Earth's systematic temperature response are better informed by making physical assumptions than by an ostensibly more empirical approach. In this section, we ask a similar question about characterizing internal variability, in settings similar to that discussed, where data are temporally correlated and limited in length. We used a Gaussian ARMA(4,1) noise model in Section~\ref{sec:UQ}; while we do not claim a physical motivation for this model, the model does make relatively strong statistical assumptions. Is it more robust to assume a low-order parametric model for the noise, as we have done, or to adopt a nonparametric approach? The answer to this question depends on the length of the data record and the nature of the internal variability.

For the purposes of this illustration, we will use not the actual temperature record but rather some simple synthetic examples. We consider several artificial, trendless time series (the true mean of the process is constant) with temporally correlated noise, and evaluate the results of testing for a linear time trend (i.e., fitting model~\eqref{eq:timeModel} and testing against the null hypothesis that $\beta=0$) using different parametric (both correctly specified and not) and nonparametric methods for estimating uncertainties. The ordinary least squares standard errors, which assume uncorrelated noise, tend to be anticonservative (too small) and time series methods are supposed to ameliorate this overconfidence, but may or may not be successful in this regard depending on the context. The illustrations are simple but our conclusions should be relevant to actual data analysis and to more informed models.

We consider a few parametric approaches common in time series analysis. The typical practice is to assume that the noise follows a low order model, such as an ARMA model. Ideally, the noise model would be chosen (as in Section~\ref{sec:UQ}) in consultation with diagnostic plots and an information criterion like the AICc that balances model fit with the number of statistical parameters. However, it is common practice to automatically select a model either solely by minimizing the AICc or another criterion, or just to assume a simple time series model, often the AR(1) model. Uncertainty in the trend parameter(s) can then be estimated in several ways. Usually, this is done assuming asymptotic normality for maximum likelihood estimators, and a test might be carried out using a $t$-statistic. If asymptotic normality is not viable, as for model~\eqref{eq:emulator}, an alternative is to resort to using a parametric bootstrap (again as we do in Section~\ref{sec:UQ}) and to perform the test using bootstrap $p$-values.

We also evaluate the perhaps most typical nonparametric method for accounting for dependence in a time series, the block bootstrap \citep{kunsch}. Here, residuals are resampled in blocks to generate bootstrap samples that retain much of the dependence structure in the original data. A popular variant is the circular block bootstrap \citep{politis}, in which blocks can be overlapping and blocks starting at the end of the time series wrap back to the beginning. The block bootstrap has been a common method in the climate and atmospheric sciences for at least two decades \citep{wilks} and has been applied previously to test for time trends (or changes thereof) in the global mean surface temperature record. For example, \cite{rajaratnam} argue that the circular block bootstrap gives better uncertainty estimates than does a parametric analysis in the setting of testing for a trend in a 16-year segment of the global mean temperature record.

While the block bootstrap works very well in some settings, the procedure is not free of assumptions. Like other variants of the nonparametric bootstrap, its justification is based on an asymptotic argument: for the block bootstrap to work well, the size of the block has to be small compared to the overall length of the data but large compared to the scale of the temporal correlation in the data. When the overall data record is short and internal variability is substantially positively correlated in time, as for the historical global mean temperature record, these dual assumptions may not both be met and we should not expect the block bootstrap to perform well.

In the following, we compare five methods (four parametric and one nonparametric) for generating nominal $p$-values testing for a significant linear time trend in trendless synthetic data: (a) a $t$-test assuming independent noise, (b) a $t$-test assuming Gaussian AR(1) noise, (c) bootstrap $p$-values from a parametric bootstrap assuming Gaussian AR(1) noise, (d) a $t$-test assuming Gaussian ARMA($p$,$q$) noise with the order chosen by minimizing AICc, and (e) bootstrap $p$-values from a circular block bootstrap. We generate synthetic data from three different time series models, but in cases (b) and (c), the assumed parametric model is always AR(1).  (Table A3 summarizes the models from which we simulate, which were chosen in part to share a similar strength of correlation with the observed global mean temperature record.) In cases (a), (b), and (d), the $t$-test degrees of freedom are approximated by $n-2$ minus the number of parameters in the noise model. The parametric models in cases (a) through (d) are all estimated via maximum likelihood. In case (e), we estimate $p$-values using a few different block lengths and show the results most favorable to the block bootstrap. 

After generating nominal $p$-values, we evaluate the performance of the different methods. Since the null hypothesis is true in this artificial setting (the synthetic series are trendless), a correct $p$-value would be uniformly distributed. Deviations from the uniform distribution would then be a sign that the procedure generating the nominal $p$-value is uncalibrated. We therefore use Q-Q plots to compare the distribution of the simulated nominal $p$-values with the theoretical uniform distribution. Uncertainties are underestimated when the nominal $p$-value quantiles are smaller than the theoretical quantiles (i.e., when the Q-Q plot is below the one-to-one line). In this situation, inferences are anticonservative and the tests using the selected method will have Type I error rates that are larger than the nominal rate.

\subsection{Parametric vs.\ nonparametric methods under a correctly specified noise model}
\begin{figure}
\begin{center}
\includegraphics[scale = 0.7]{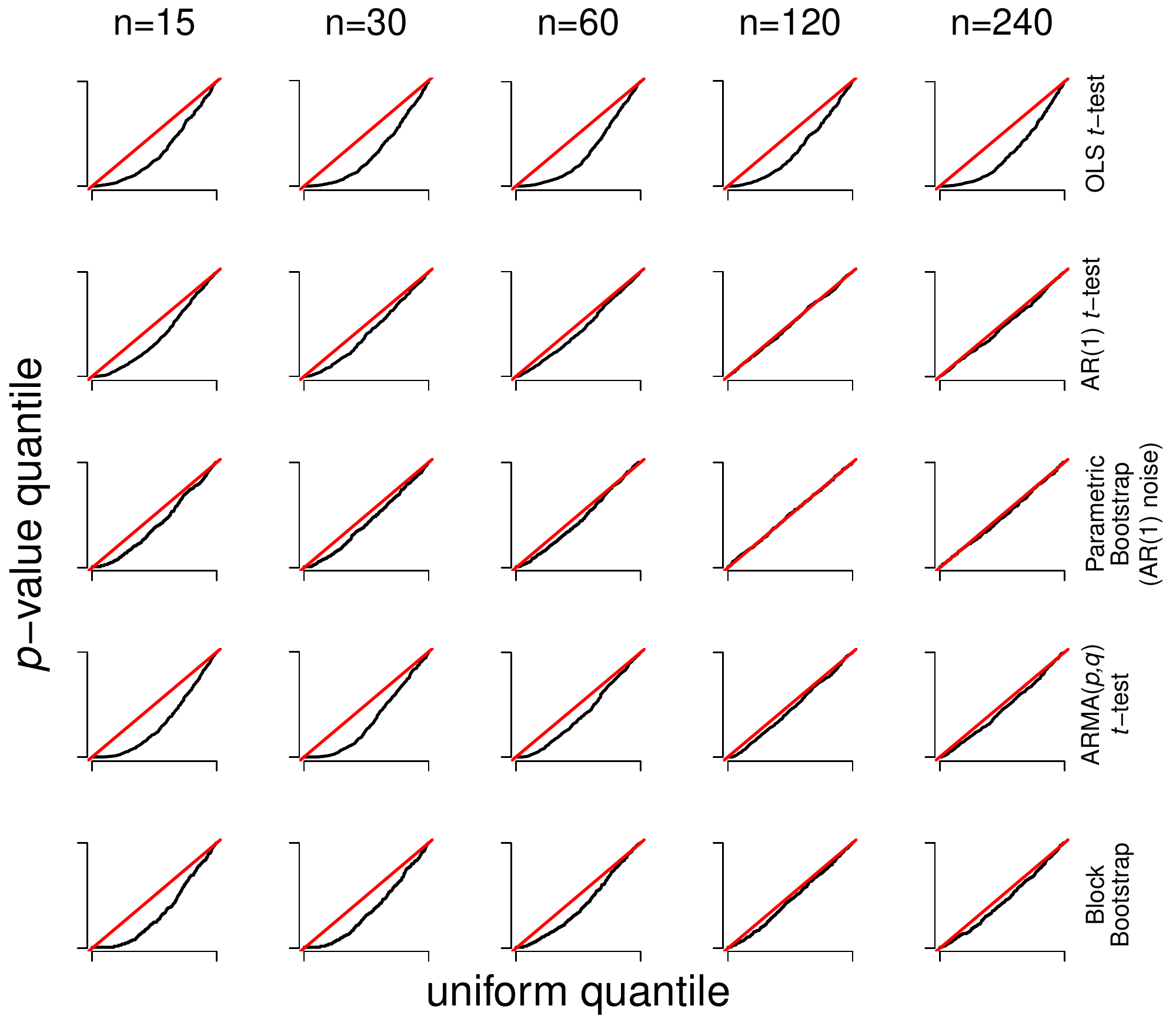}
\end{center}
\caption{Quantile-quantile plots comparing the distribution of nominal $p$-values to the theoretical uniform distribution. Simulations are of mean zero, Gaussian AR(1) time series and $p$-values correspond to a two-sided test for a linear time trend. The length of the time series is given above each column. In the first row, the $p$-values are from the OLS $t$-test assuming independent noise; the second row is the $t$-test assuming AR(1) noise; the third row uses a parametric bootstrap (again assuming AR(1) noise); the fourth row uses a $t$-test assuming ARMA($p$,$q$) noise with the order of the model selected by AICc; and the last row uses a circular block bootstrap with block size chosen to be favorable to this method. The $p$-values calculated assuming the correct parametric model appear approximately correct for modest sample sizes and always outperform both blind selection by AICc and the block bootstrap. The latter two methods can be worse than assuming independence when sample sizes are very small.}
\label{fig:QQAR1}
\end{figure}

We first compare the performance of the five methods for generating nominal $p$-values in the setting where the assumed AR(1) model is correctly specified (Figure~\ref{fig:QQAR1}).  Unsurprisingly, pre-specified parametric time series methods give reasonably calibrated inferences when the parametric model is correctly specified (Figure~\ref{fig:QQAR1}, rows 2 and 3), although maximum likelihood gives somewhat anticonservative estimates of uncertainty with small sample sizes. The anticonservative bias of the maximum likelihood estimator can be reduced by instead using restricted maximum likelihood (REML) (e.g.,~\cite{mcgilchrist}) (see Appendix A5), but we focus here on the performance of maximum likelihood because it is more usually the procedure employed. It is also unsurprising that $p$-values under assumed independence are quite uncalibrated and anticonservative in this setting (Figure~\ref{fig:QQAR1}, top row), because standard errors are underestimated when positive temporal correlation is ignored.

It may, on the other hand, be surprising that automatically chosen parametric methods and nonparametric methods (blind selection via AICc and the block bootstrap, Figure~\ref{fig:QQAR1}, rows 4 and 5) can perform \textit{even more poorly than assuming independence} if the sample size is very small. The AICc does improve on the AIC (not shown) by attempting to account for small sample biases, but still performs poorly in very small samples. The (nonparametric) block bootstrap is the worst-performing method at the smallest sample sizes while ostensibly based on the weakest assumptions. For either of these methods to perform comparably to the pre-specified parametric methods, sample sizes must be quite large, indeed in this illustration larger than the available global mean temperature record. This should serve as a warning against using these methods for time series of modest length.

In actual practice, it can be advantageous, as we already discussed, to choose a noise model not automatically but in consultation with diagnostics (such as by comparing theoretical spectral densities with the empirical periodogram). In Section~\ref{sec:UQ}, we chose a noise model with consideration for the model's representation of low-frequency variability. In that example, the chosen model did minimize the AICc, but gave more conservative inferences than the AR(1) model (which was chosen by BIC). The illustration above shows that this behavior is not generically true and that for small sample sizes it is dangerous to blindly select models by minimizing an information criterion alone.

\subsection{Parametric vs.\ nonparametric methods under a misspecified model}
\label{subsec:misspecified}
\begin{figure}
\begin{center}
\includegraphics[scale = 0.5]{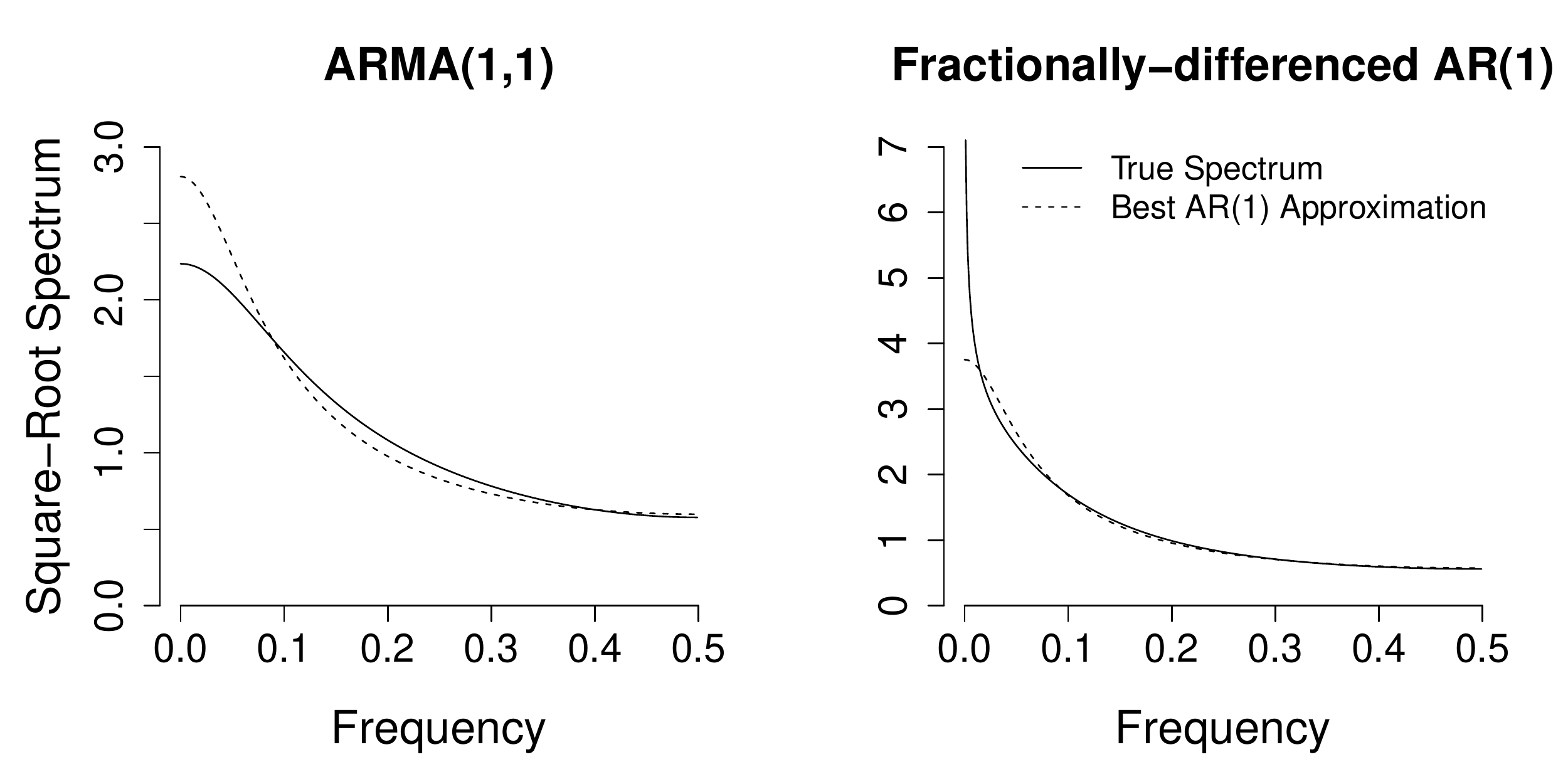}
\end{center}
\caption{The power spectra associated with the two models from which we generate synthetic time series in Section~\ref{subsec:misspecified}, along with spectra of the best AR(1) approximations (in Kullback-Leibler divergence) to both models; see Table A3 for noise model parameters. The corresponding illustrations of the performance of the various time series methods are shown in Figures~\ref{fig:QQARMA11} and~\ref{fig:QQFAR}. The AR(1) model will tend to overestimate low-frequency variability in the first case and underestimate it in the second.}
\label{fig:SpecDenSimulPlot}
\end{figure}

\begin{figure}
\begin{center}
\includegraphics[scale = 0.7]{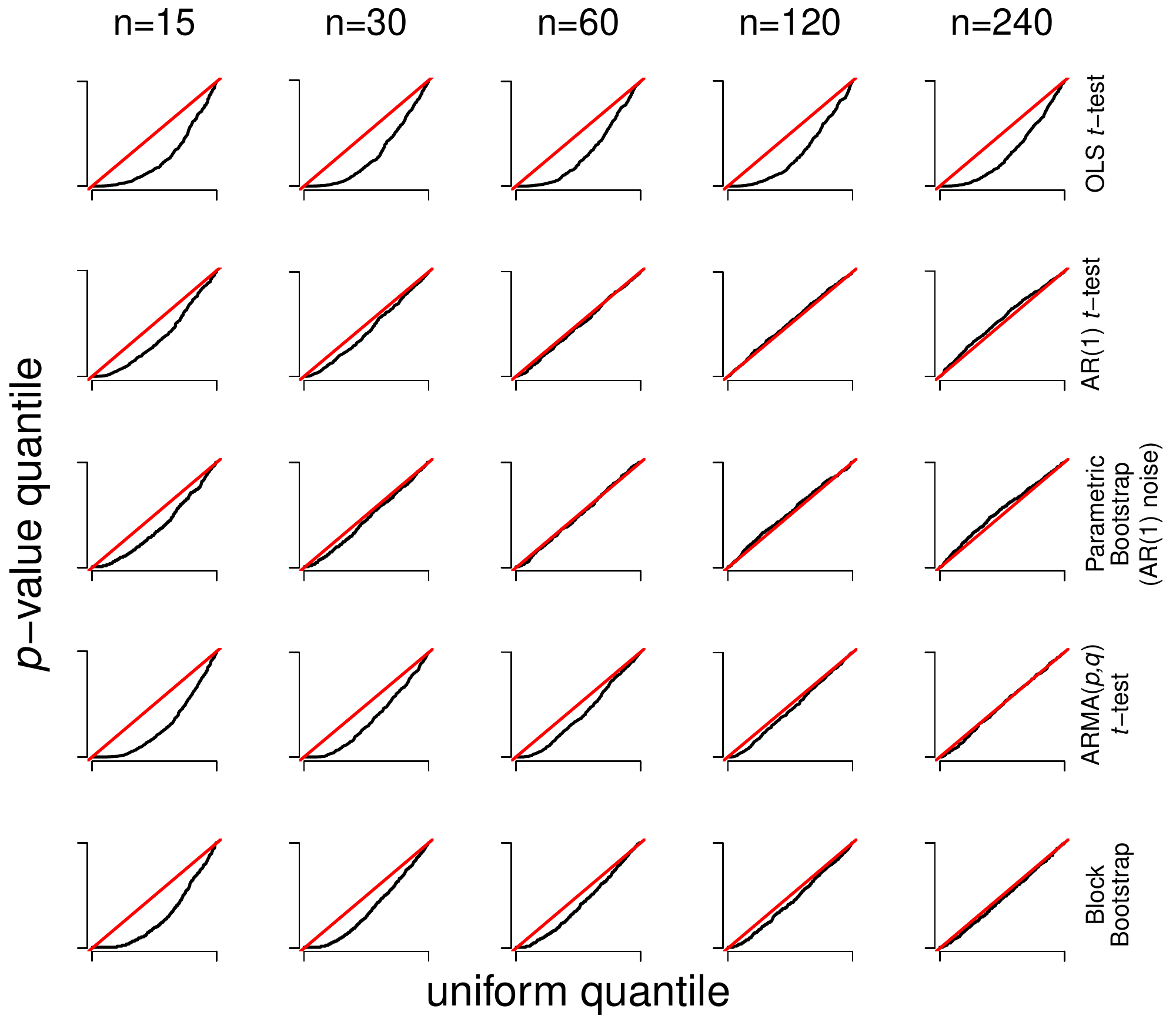}
\end{center}
\caption{Same as Figure~\ref{fig:QQAR1} but with simulations from an ARMA(1,1) model. The $p$-values calculated by incorrectly assuming an AR(1) model are increasingly conservative in larger sample sizes in this setting. Both blind selection via AICc and the block bootstrap are anticonservative for small sample sizes but improve as the sample size increases.}
\label{fig:QQARMA11}
\end{figure}

\begin{figure}
\begin{center}
\includegraphics[scale = 0.7]{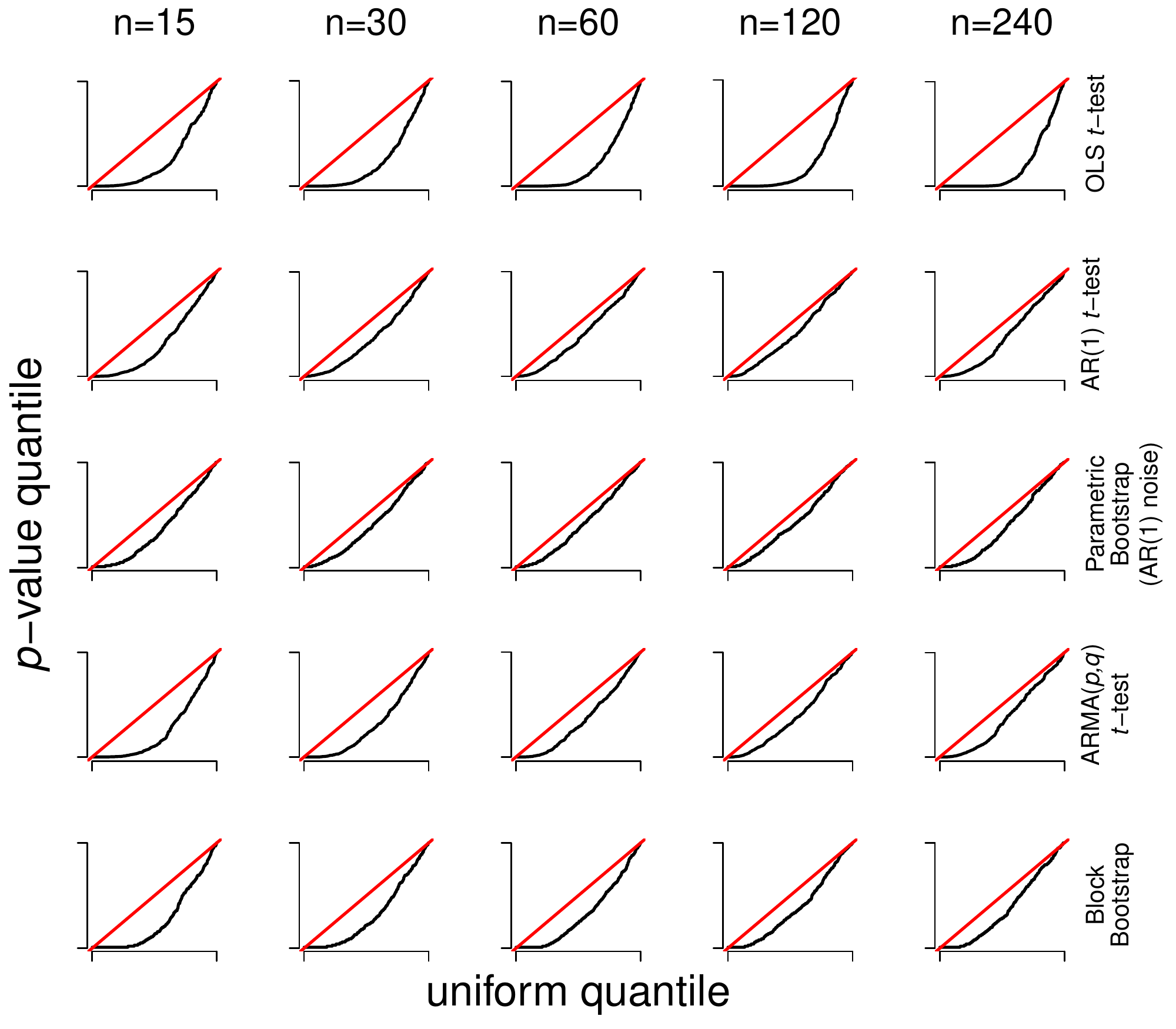}
\end{center}
\caption{Same as Figure~\ref{fig:QQAR1} but with simulations from a fractional AR(1) model.  While none of the methods perform very well here, the incorrectly specified parametric methods are better, especially in smaller samples.}
\label{fig:QQFAR}
\end{figure}

The comparisons in the previous section were too favorable to the pre-specified parametric methods because the order of the specified noise model (an AR(1) model) was known to be correct. Now we compare these methods when the assumed noise model is misspecified. The performance of misspecified methods will depend in particular on how the misspecified model represents low- versus high-frequency variations in the noise process. Models that underestimate low-frequency variability will tend to be anticonservative for estimating uncertainties in smooth trends, whereas those that over-estimate low-frequency variability will tend to be conservative. We therefore repeat the illustrations in the previous section generating the synthetic time series from two different noise models (but still using the pre-specified AR(1) model to generate nominal $p$-values). The two models are chosen so that their best AR(1) approximations either under- or over-represent low-frequency variability. Figure~\ref{fig:SpecDenSimulPlot} shows the spectra corresponding to these two noise models and the best AR(1) approximation to each, and Figures~\ref{fig:QQARMA11} and \ref{fig:QQFAR} show Q-Q plots corresponding to how the various time series methods perform in these two settings. See Table A3 for the model parameters and Section A3 for a comparison under two additional models, including when the true model is the ARMA(4,1) model we assume in our main analysis of the global mean temperature record.

First, we consider an ARMA(1,1) process whose best AR(1) approximation over-represents low-frequency variability (Figure~\ref{fig:QQARMA11}). The results are similar to those when the model was correctly specified, except that at the largest sample sizes the pre-specified parametric methods slightly overestimate uncertainties, for the reasons discussed above. As before, both blind selection via AICc and the block bootstrap perform well for large sample sizes but very poorly for the smallest sample sizes.

Second, we consider a fractionally integrated AR(1) process; because this is a long-memory process, the best AR(1) approximation (and indeed any ARMA model) will severely under-represent low-frequency variability (Figure~\ref{fig:QQFAR}). All of the methods struggle in this setting and produce anticonservative estimates, but the pre-specified parametric methods still typically perform better than the ostensibly more flexible methods.

These results confirm that approaches to representing noise that appear to weaken assumptions are not guaranteed to outperform even misspecified parametric models. Misspecified parametric models are most dangerous when low-frequency variability is under-represented, but methods like the block bootstrap will also have the most trouble when low-frequency variability is strong because very long blocks will be required to adequately capture the scale of dependence in the data. While it is  crucial for the data analyst to scrutinize any assumed parametric model, we believe that in many settings when the time series is not very long relative to the scale of correlation, one will be better served by carefully choosing a low-order parametric model rather than resorting to nonparametric methods.

Note that this illustration uses synthetic simulations that are relatively strongly correlated in time, a feature of the global mean temperature record. Nonparametric methods can work better than illustrated here in settings where correlations are weaker. For example, \textit{local} (rather than global) temperatures, at least over land, tend to be more weakly correlated in time. In general, it is useful to evaluate the performance of of statistical methods with simulations that share characteristics with the relevant real data.

\section{Discussion}
\label{sec:discussion}
\label{sec:discussion}
We have sought to show here that targeted parametric modeling of global mean temperature trends and internal variability can provide more informative and accurate analyses of the global mean temperature record than can more empirical methods. Since all analyses involve assumptions, it is important to consider the role that assumptions play in resulting conclusions. In the setting of analyzing historical global mean temperatures, where the data record is relatively short and temporal correlation is relatively strong, ostensibly more empirical methods can fail to distinguish between systematic trends and internal variability, and can give seriously uncalibrated estimates of uncertainty. While linear-in-time models can be used for some purposes when applied to moderately narrow time frames (and with careful uncertainty quantification), the demonstrations shown here suggest that they do not have an intrinsic advantage over more targeted analyses. Targeted analyses can be used over longer time frames -- allowing for better estimates of both trends and noise characteristics -- and can address a broader range of questions within a single framework.

The model we use in our analysis provides insights about the information contained in the historical temperature record relevant to both shorter-term and longer-term trend projections. The limited historical record of global mean temperature can provide information about shorter-term trends but unsurprisingly cannot constrain long-term projections very well. The past 136 years of temperatures simply do not alone contain the relevant information about equilibration timescales that would be required to constrain long-term projections, especially when aggregated to a single global value. (Use of spatially disaggregated data may provide additional information.) The distinction between uncertainties in shorter-term and longer-term projections, itself not easily made using a time trend model, serves to further illustrate that while the historical data record is an important source of information, it alone cannot be expected to answer the most important questions about climate change without bringing more scientific information to bear on the problem.

We believe that our discussion is illustrative of broader issues that arise in applied statistical practice, and will have particular relevance to problems involving trend estimation in the presence of temporally correlated data and in relatively data-limited settings, common in climate applications. We suspect that many applied statisticians have personally felt the tension between targeted modeling on the one hand and more empirical analyses on the other. One lucid discussion of the broader issues surrounding this tension can be found in the discussion of model formulation in~\cite{cox}. Empirical approaches have very successfully generated new insights and predictions in important areas, but there is a wide range of scientific problems where these approaches do not perform well and where targeted, domain-specific modeling is required. Statisticians can bring important insights to scientific problems; a crucial role for the statistician is to consider the modeling choices that will be both the most illuminating and the most reliable given the scientific questions and data at hand.

\section{Data availability}
The NASA GISS Land-Ocean Temperature index is updated periodically; the data we analyze was accessed on the date 2016-02-03. The current version is available at http://data.giss.nasa.gov/gistemp/. The HADCRUT4 data, used in Section 4.5, is available at http://www.metoffice.gov.uk/hadobs/hadcrut4/data/current/download.html.

Historical radiative forcings until 2011 are available in ~\cite{AR5} Table AII.1.2. Forcings corresponding to the RCP 8.5 scenario can be found at http://tntcat.iiasa.ac.at/RcpDb. NOAA CO$_2$ concentrations are available at\\
\noindent ftp://aftp.cmdl.noaa.gov/products/trends/co2/co2\_annmean\_gl.txt

\section*{Acknowledgements}
The authors thank Jonah Bloch-Johnson, Malte Jansen, Cristian Proistosescu, and Kate Marvel for helpful conversations and comments related to parts of this work. We additionally thank the reviewers of this paper, whose suggestions led to a number of improvements. This work was supported in part by STATMOS, the Research Network for Statistical Methods for Atmospheric and Oceanic Sciences (NSF-DMS awards 1106862, 1106974 and 1107046), and RDCEP, the University of Chicago Center for Robust Decisionmaking in Climate and Energy Policy (NSF Grant SES-0951576). We thank NASA GISS, NOAA, the Hadley Centre, the IPCC, and IIASA for the use of their publicly available data. We acknowledge the University of Chicago Research Computing Center, whose resources were used in the completion of this work.

\renewcommand{\thesection}{A\arabic{section}}   
\renewcommand{\thetable}{A\arabic{table}}   
\renewcommand{\thefigure}{A\arabic{figure}}
\renewcommand{\theequation}{A\arabic{equation}}
\setcounter{section}{0}
\setcounter{table}{0}
\setcounter{figure}{0}
\setcounter{equation}{0}

\section{Implications for the transient climate response and equilibrium sensitivity}
\label{sec:TCRandECS}
In this paper, we use the historical temperature record to estimate the trend model~\eqref{eq:emulator} that treats both the sensitivity parameter, $\lambda_A$, and the timescale of response parameter, $\rho_A$, as unknown. In this section, we compare our results to those from more commonly used methods of inferring the climate response from projections in climate models and of using the historical temperature record to estimate a climate sensitivity parameter.

Model~\eqref{eq:emulator} is justifiable over the relatively short (centennial) timescales over which the \textit{linear forcing-feedback model} is applicable. The linear forcing-feedback model says that the mean temperature response to an instantaneous forcing $F$ behaves like $L \Delta T(t) \approx F - Q(t)$, where $\Delta T(t)$ is the change in mean temperature at time $t$ from the baseline state, $Q(t)$ is the heat uptake (in W/m$^2$) at time $t$, and $L$ is a climate response parameter (with units W/m$^2$/K). (In the context of this model, our sensitivity parameter is then $\lambda_A = F_{2\times} / L$.) Our method essentially infers $Q(t)$ by assuming that it decays exponentially until reaching zero again at equilibrium. The linear forcing-feedback model does not hold for all time, however, in part because both feedbacks and rates of warming vary spatially. In models, this typically results in additional warming over the millennial timescales on which the deep ocean mixes (e.g.,~\cite{winton,armour,rose} and others). Additionally, climate feedbacks -- and therefore climate sensitivity -- may be state-dependent; this effect also typically amplifies global mean temperature rise~\citep{blochjohnson}. Collectively, this suggests that our sensitivity parameter $\lambda_A$ will be smaller than the equilibrium climate sensitivity that measures the final, equilibrium temperature response.

We can compare the results of our model to reported results from GCMs by estimating the \textit{transient climate response} (TCR), a popular metric of the short-term temperature response to forcing in climate models. The TCR is defined as the change in mean temperature after 70 years of a CO$_2$ concentration scenario that increases by 1\% per year (so doubles after 70 years). In a multi-model comparison  of these centennial-scale projections,~\cite{AR5} reports a 66\% ``likely'' interval for the TCR of 1.0 to 2.5$^\circ$C. Our results from fitting model~\eqref{eq:emulator} to the historical data are fairly consistent with this: our estimate of the TCR is 1.7$^\circ$C, with the 2.5-97.5th bootstrap percentile interval of 1.2 to 1.9$^\circ$C.  These intervals are not strictly comparable because the IPCC's subjectively combines information from multiple lines of evidence. That said, our interval is shorter in part because it does not account for uncertainties in historical radiative forcings from anthropogenic aerosols. If we repeat the exercise of Section 4.5 (scaling the past radiative forcing trajectory to approximate the upper and lower bounds for forcing accounting for aerosol uncertainties), our central estimate of the TCR would be about 1.2$^\circ$C if anthropogenic aerosols were on the high end and about 3.4$^\circ$C if on the low end.

We can also compare our estimate of $\lambda_A$ to prior estimates of a sensitivity parameter that also use the historical temperature record. The most typical approach in the literature shares some commonalities with our method, beginning with the same linear forcing-feedback model that implicitly underlies our analysis, but estimating a sensitivity parameter by using an additional observational estimate of global heat uptake. That is, studies use estimates of changes in forcing, $F'$, heat uptake $Q'$, and historical temperature change, $\Delta T'$, between a base and a final period to compute $\hat{L} = (F'-Q')/\Delta T'$ and therefore $\hat{\lambda} = F_{2\times} \Delta T'/ (F'- Q')$. Studies using this method include~\cite{gregory2002},~\cite{otto},~\cite{masters}, and~\cite{lewis}. The resulting sensitivity parameter estimate should be similar to our $\lambda_A$. Table~\ref{tab:ECS} shows the results from these analyses; the central estimates are indeed similar to our estimate of $\lambda_A$. The uncertainty ranges given in these studies, however, also tend to be slightly larger than the intervals we give for $\lambda_A$, again because these authors attempt to account for radiative forcing uncertainty in their analysis.

\begin{table}
\caption{Comparison of estimates of a sensitivity parameter from studies that use observational data and a simple energy balance approach. The large best (median) estimate from~\cite{gregory2002} is due to a very fat right tail in their analysis; the mode of their distribution is 2.1$^\circ$C per doubling. The estimates given in these studies are similar to our estimates of $\lambda_A$ (which should be smaller than the equilibrium sensitivity).}
\label{tab:ECS}
\begin{center}
\begin{tabular}{c ll | c c}
&  & \multirow{2}{*}{Study} & Best estimate & 90\% Interval \\
& & & ($^\circ$C per doubling) & ($^\circ$C per doubling) \\
\hline
\textit{Energy balance} & \rdelim\{{4}{3mm} & \cite{gregory2002} & 6.1 & $>$1.6 \\
\textit{model using} &  &\cite{otto} & 2.0 & 1.2-3.9 \\ 
\textit{temperatures, forcing} & & \cite{masters} & 2.0 & 1.2-5.2 \\
\textit{and heat uptake} & & \cite{lewis} & 1.6 & 1.1-4.1 \\ 
 \\ 
 \hline
 \\
\textit{Trend model~\eqref{eq:emulator}} & & this work & 1.8 & 1.5-3.0 \\
\hline
\end{tabular}
\end{center}
\end{table}

A distinguishing feature of this other, common approach is that it includes data about historical heat uptake, in addition to temperature and radiative forcing data. Ocean heat content is an additional, albeit uncertain, source of information that may improve estimates.  On the other hand, since these methods do not involve an explicit trend model and require averaging the inputs over decadal or longer timespans, they cannot use the historical temperature record to estimate internal temperature variability. Most studies therefore estimate internal variability using climate model output, but climate models do not perfectly realistically represent even present-day variability in global annual mean temperature. 

By contrast, an advantage of our approach is that it allows one to use the historical data to understand internal variability. Additionally, our approach allows one to answer questions about both historical trends and longer-term projections in the framework of one statistical model, whereas the approaches discussed above do not allow one to infer trends in increasing-in-time forcing scenarios. A disadvantage of our approach is that, as discussed above, we rely on the historical global mean temperature record to estimate the ``equilibration'' timescales ($\rho_A$ and $\rho_N$), but the data contain little information about these quantities.

Regardless of the different advantages and disadvantages just discussed, both approaches to using the historical temperature record give similar results concerning the sensitivity parameter, and uncertainties in this parameter remain high. This demonstrates the limitations of the information content of the historical global mean temperature record alone for estimating longer-term projections of mean temperature changes. As noted in Section~\ref{sec:discussion}, spatially disaggregated data may contain more information.

\section{Coefficient estimates for noise model}
\label{sec:coefs}
Table~\ref{tab:ARMAParams} gives information about the ARMA(4,1) and AR(1) models fit to the residuals of model~\eqref{eq:emulator} used in Section 4. The ARMA(4,1) model give more conservative inferences about the systematic trend, and is the model we adopt in Section 4 (see Figure~\ref{fig:resSpectrum}). 

\begin{table}
\small
\begin{center}
\begin{tabular}{l | c c c c  c  c c c}
\multirow{2}{*}{Model} & \multicolumn{4}{c }{Autoregressive} &  Moving average & Innovation & & \\
& \multicolumn{4}{c}{parameters} & parameter  & standard & & \\
 & $\phi_1$ & $\phi_2$ & $\phi_3$ & $\phi_4$ & $\theta_1$ & deviation ($^\circ$C) & AICc & BIC \\
 \hline
\multirow{2}{*}{ARMA(4,1)} & -0.29 & 0.36 & 0.05 & 0.24 & 0.80 & 0.09 & -258 & -242\\
& (0.19)  & (0.12) & (0.09) & (0.09) & (0.19) &  &  & \\
\multirow{2}{*}{AR(1)} & 0.52 & -- & -- & -- & -- & 0.09  & -257 & -252\\
 & (0.07) & & & & &  &  & \\
\end{tabular}
\end{center}
\caption{Coefficient estimates (with standard errors in parentheses), innovation standard deviations, and AICc and BIC associated with ARMA(4,1) and AR(1) fits to the residuals from model~\eqref{eq:emulator}. These two models minimize AICc and BIC, respectively. We use the ARMA(4,1) model because this model appears to better represent the low-frequency variation in the residuals (see Figure~\ref{fig:resSpectrum}), which is crucial for inferences about trends.}
\label{tab:ARMAParams}
\end{table}

\section{Performance of misspecified methods under AR(2) and ARMA(4,1) simulations}
\label{sec:AR2Comp}
\begin{figure}
\begin{center}
\includegraphics[scale = 0.5]{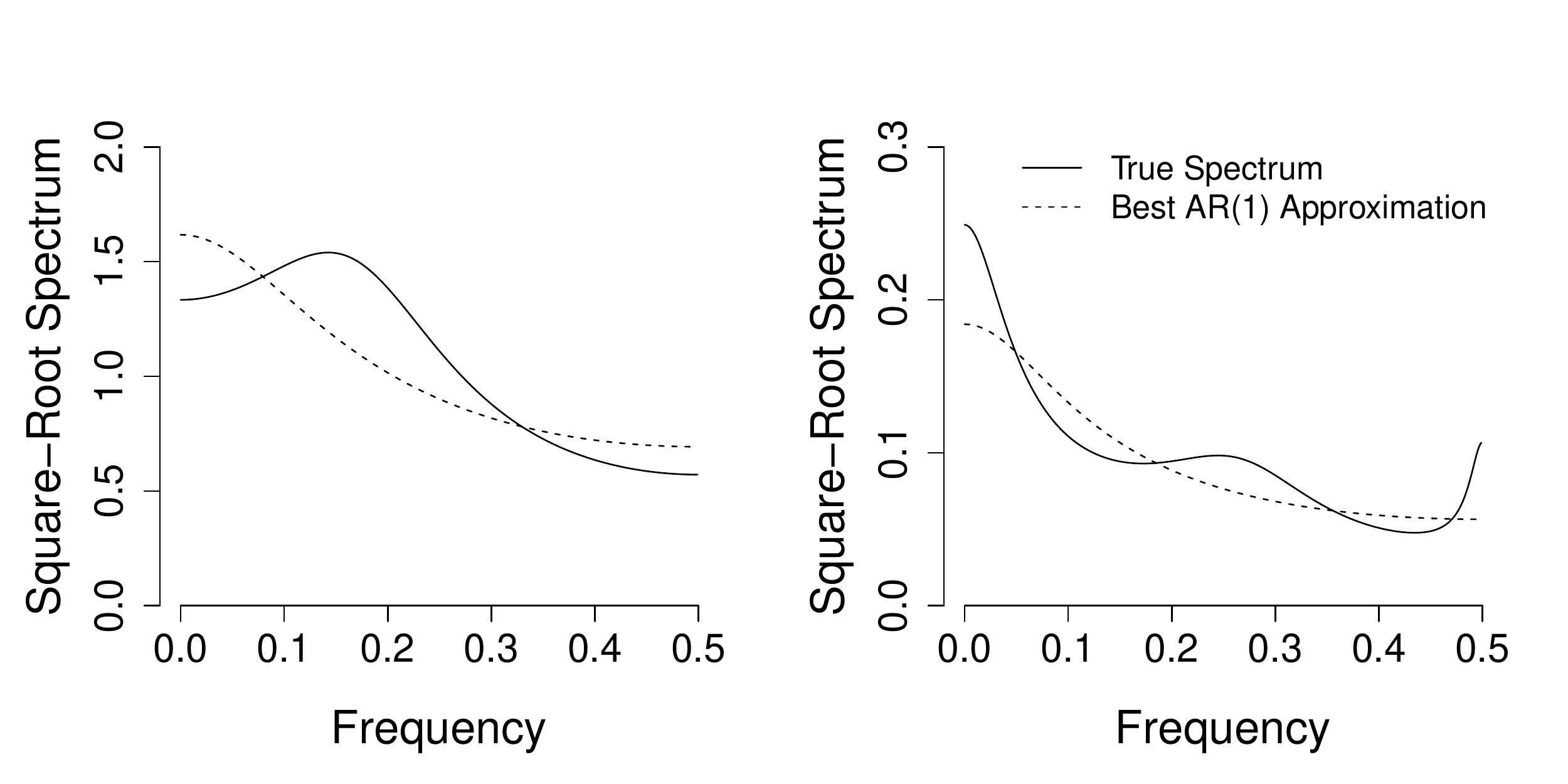}
\end{center}
\caption{Analogous to Figure~\ref{fig:SpecDenSimulPlot} but for the AR(2) (left) and ARMA(4,1) (right) models considered in Section~\ref{sec:AR2Comp}. See Table~\ref{tab:parameters} for noise model parameters.}
\label{fig:SpecDenSimulPlotAR2ARMA41}
\end{figure}

In Section~\ref{subsec:misspecified}, we compared the five methods for generating nominal $p$-values in two settings where the pre-specified AR(1) model was incorrect, where the AR(1) model either over- or under-represented low-frequency variability. Here we present two more comparisons. For these, Table~\ref{tab:parameters} gives the model parameters and Figure~\ref{fig:SpecDenSimulPlotAR2ARMA41} shows the two true spectra and the corresponding best AR(1) approximations. The first comparison is to an AR(2) noise model under which the AR(1) approximation over-represents variability at the lowest frequencies but under-represents variability at intermediate frequencies (Figure~\ref{fig:SpecDenSimulPlotAR2ARMA41}, left). The second comparison is to the ARMA(4,1) model used in our main analysis, under which the AR(1) approximation under-represents variability at the lowest frequencies (Figure~\ref{fig:SpecDenSimulPlotAR2ARMA41}, right).

\begin{figure}
\begin{center}
\includegraphics[scale = 0.5]{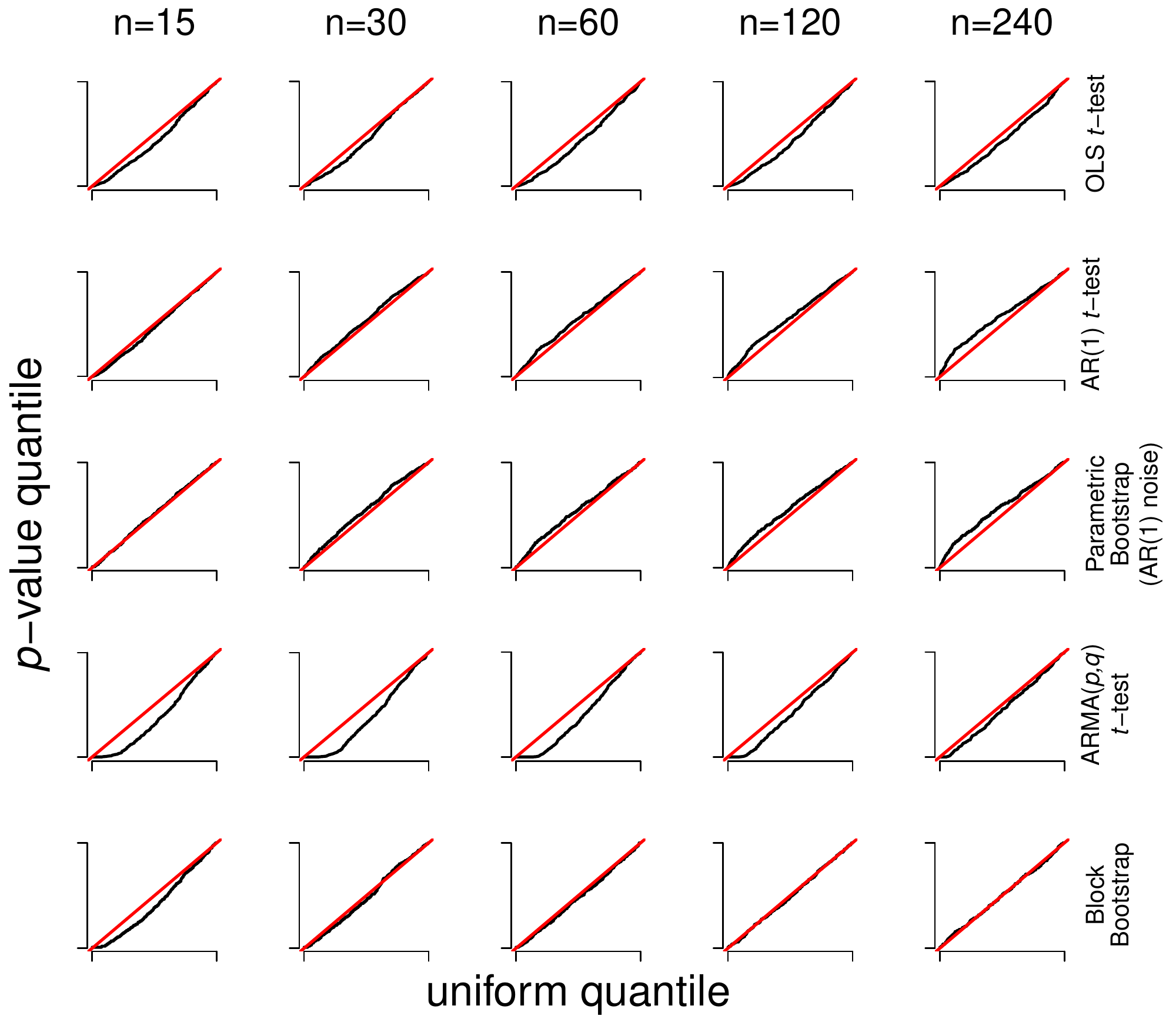}
\end{center}
\caption{Same as Figure~\ref{fig:QQAR1} but with simulations from an AR(2) model. The relative results are similar to the setting of Figure~\ref{fig:QQARMA11} but all methods perform better here.}
\label{fig:QQAR2}
\end{figure}

First, we show results for the AR(2) true noise model model (Figure~\ref{fig:QQAR2}). In this setting, all methods perform better than in the setting of Figure~\ref{fig:QQARMA11}, but the relative performance of the different methods is largely the same as there, except that selection by AICc now performs even more poorly than the block bootstrap at the smallest sample sizes. 

\begin{figure}
\begin{center}
\includegraphics[scale = 0.5]{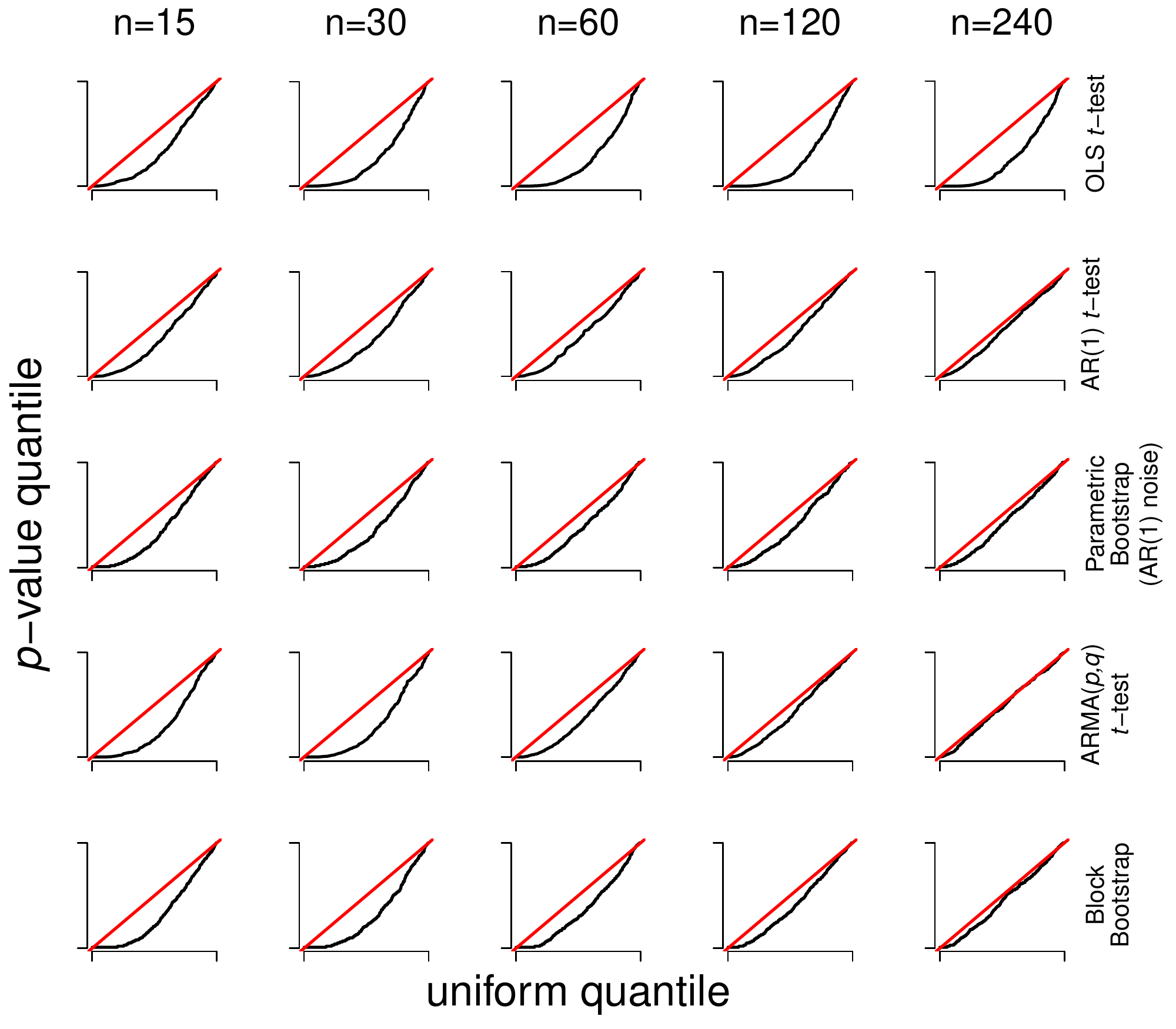}
\end{center}
\caption{Same as Figure~\ref{fig:QQAR1} but with simulations from a the ARMA(4,1) model used as the noise model in our main analysis. The relative results are similar to but less extreme than those in the setting of Figure~\ref{fig:QQFAR}.}
\label{fig:QQARMA41}
\end{figure}

Second, we show results for the ARMA(4,1) model used in our main analysis (Figure~\ref{fig:QQARMA41}). In this setting, the results are similar in nature to but less extreme than those from the fractionally differenced AR(1) model shown in Figure~\ref{fig:QQFAR}. All methods perform poorly but the parametric methods tend to perform better, especially when the time series length is short. 

The behavior from both of these simulations again serves to emphasize that when making inferences about smooth trends, it is most crucial to represent low-frequency variability well.

\section{Model coefficients for simulations in Sections~\ref{sec:ParaVsNonpara} and~\ref{sec:AR2Comp}}
\label{sec:coefsTSMethods}
In Sections~\ref{sec:ParaVsNonpara} and~\ref{sec:AR2Comp}, we generate synthetic, mean zero time series with different correlation structures. The models that we simulate from are summarized in Table~\ref{tab:parameters}. The general form for an autoregressive fractionally integrated moving average  model (ARFIMA) of order $(p,d,q)$ (of which all the simulated models are special cases) is
$$\left(1-\sum_{k=1}^{p}\phi_k B^k \right) (1-B)^d Y_t = (1+\sum_{k=1}^{q} \theta_k B^k) \epsilon(t),$$
where $Y_t$ is the time series at time $t$, the $\phi$'s are the AR parameters, the $\theta$'s are the MA parameters, $d$ is the (fractional) differencing parameter, $B$ is the backshift operator (i.e., $B^kY_t = Y_{t-k}$), and $\epsilon(t)$ are uncorrelated innovations with constant variance. (The convention is that the acronym is shortened to account for parameters that are set to zero, so for example an ARFIMA(1,0,0) model is called an AR(1) model.)
\begin{table}
\small
\begin{center}
\begin{tabular}{c | l c c c}
\multirow{2}{*}{Figure} & \multirow{2}{*}{Name} & AR & MA & Differencing\\
& & parameter(s) & parameter & parameter \\
\hline
\ref{fig:QQAR1} & AR(1) & 0.5 & 0 & 0 \\
\ref{fig:QQARMA11} & ARMA(1,1) & 0.5 & 0.25  & 0\\
\ref{fig:QQFAR} & fractional AR(1) & 0.5 & 0 & 0.25\\ 
\ref{fig:QQAR2} & AR(2) & (0.5,-0.25) & 0 & 0 \\
\ref{fig:QQARMA41} & ARMA(4,1) & (-0.29, 0.36, 0.05, 0.24) & 0.80 & 0 \\
\end{tabular}
\end{center}
\caption{Parameters in models from which we simulate in Sections~\ref{sec:ParaVsNonpara} and~\ref{sec:AR2Comp}. }
\label{tab:parameters}
\end{table}

\section{Performance of maximum likelihood vs. REML}
\label{sec:REMLvsML}

\begin{figure}
\begin{center}
\includegraphics[scale = 0.45]{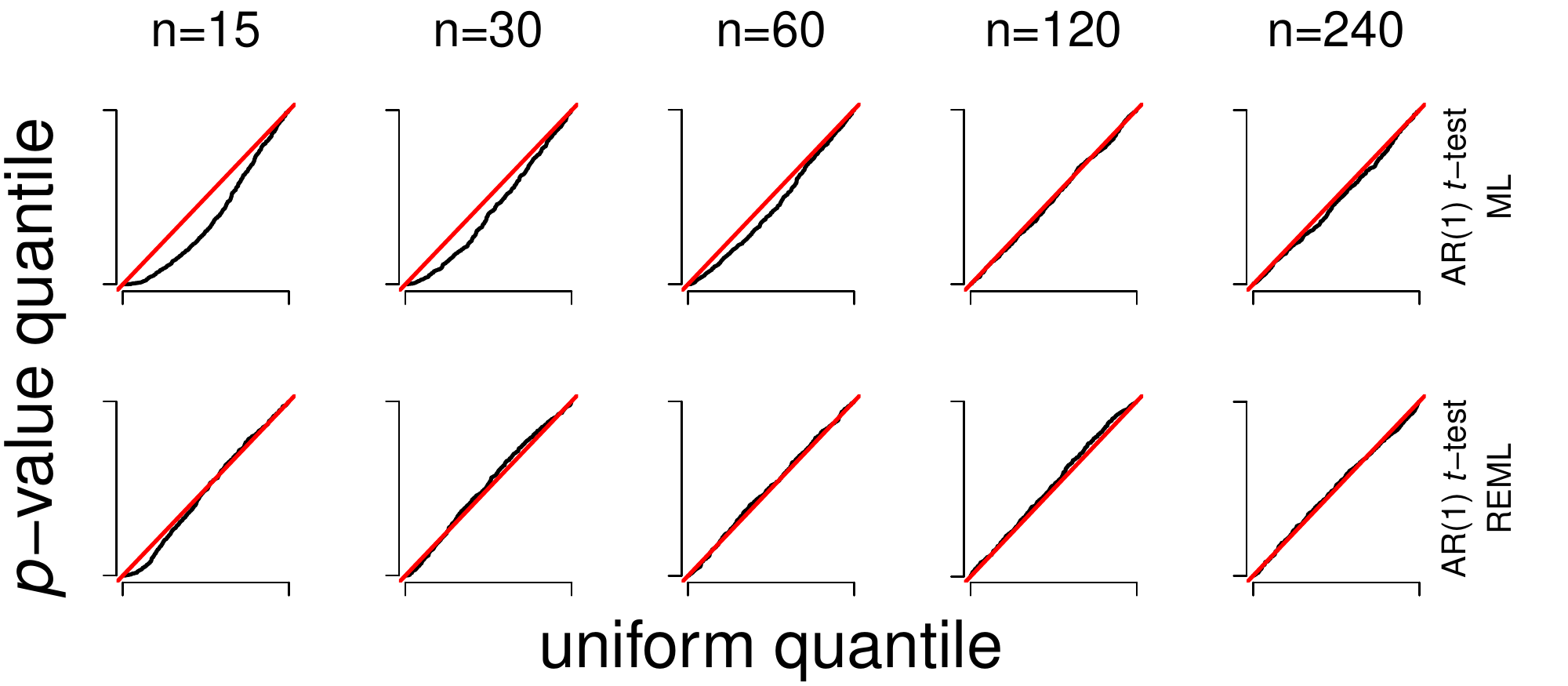}
\end{center}
\caption{Comparison of maximum likelihood and REML in the same context as Figure~\ref{fig:QQAR1}. The first row is the same as the second row of Figure~\ref{fig:QQAR1}. In the second row here, the estimation is instead done using REML. The REML standard errors give better calibrated inferences in small sample sizes.}
\label{fig:MLvsREML}
\end{figure}

In Section 5, parametric inference was done using maximum likelihood estimators. As shown there, the MLE for covariance parameters can give anticonservative estimates of standard errors for trend parameters in small sample sizes. This problem can be substantially ameliorated using restricted maximum likelihood (REML) instead. Figure~\ref{fig:MLvsREML} repeats the tests in Section 5.1 (where both the true and assumed models are AR(1)) and compares the performance of maximum likelihood and REML. For larger sample sizes, the two procedures are comparable, but in small sample sizes REML is much better calibrated (although still slightly anticonservative in the smallest sample sizes).

\bibliographystyle{chicago}
\bibliography{bibNotes}

\end{document}